\documentclass[journal,10pt]{IEEEtran}
\usepackage{balance}
\usepackage{flushend}
\usepackage[T1]{fontenc}
\usepackage{cite}
\usepackage{amsmath,graphicx}
\usepackage{amssymb}
\usepackage{algpseudocode}
\usepackage{amsfonts}
\usepackage{subfigure}
\usepackage{fancyhdr}  %
\usepackage{cases}
\usepackage{extarrows}
\usepackage{algorithm}
\usepackage{multirow,tabularx}
\usepackage{mathtools}
\usepackage{soul,color,xcolor}
\usepackage[english]{babel}
\usepackage{caption}
\usepackage{autobreak}
\usepackage{bm}
\usepackage{verbatim}
\usepackage{tikz}
\usepackage{float}  

\makeatletter
\renewcommand{\maketag@@@}[1]{\hbox{\m@th\normalsize\normalfont#1}}%
\makeatother
\newtheorem{theorem}{\textbf{Theorem}}
\newtheorem{lemma}{Lemma}

\ifCLASSINFOpdf
\else
\fi
\begin{document}
\title{Hashing Beam Training for \textcolor{black}{Integrated Ground-Air-Space Wireless Networks}}
\author{Yuan Xu, Chongwen Huang,~\IEEEmembership{Member,~IEEE}, Li Wei, Zhaohui Yang, Ahmed Al Hammadi,~\IEEEmembership{Member,~IEEE}, \\
Jun Yang, Zhaoyang Zhang,~\IEEEmembership{Senior Member,~IEEE}, Chau Yuen,~\IEEEmembership{Fellow,~IEEE}, and \\
Mérouane Debbah,~\IEEEmembership{Fellow,~IEEE}\vspace{-4mm}
\thanks{The work was supported by the China National Key R\&D Program under Grant 2021YFA1000500 and 2023YFB2904800, National Natural Science Foundation of China under Grant 62331023, 62101492, 62394292 and U20A20158, Zhejiang Provincial Natural Science Foundation of China under Grant LR22F010002, Zhejiang Provincial Science and Technology Plan Project under Grant 2024C01033, and Zhejiang University Global Partnership Fund, and Singapore Ministry of Education (MOE) Academic Research Fund Tier 2 MOE-T2EP50220-0019.\\
\indent Y. Xu and C. Huang are with College of Information Science and Electronic Engineering, Zhejiang University, Hangzhou 310027, China, with the State Key Laboratory of Integrated Service Networks, Xidian University, Xi’an 710071, China, and Zhejiang Provincial Key Laboratory of Info. Proc., Commun. \& Netw. (IPCAN), Hangzhou 310027, China (E-mails: \{yuan\_xu, chongwenhuang\}@zju.edu.cn).\\
\indent Z. Yang and Z. Zhang are with the College of Information Science and Electronic Engineering, Zhejiang University, Hangzhou 310027, China (E-mails: \{zhaohui\_yang, ning\_ming\}@zju.edu.cn).\\
\indent L. Wei and C. Yuen are with the School of Electrical and Electronics Engineering, Nanyang Technological University, Singapore 639798 (E-mails: l\_wei@ntu.edu.sg, chau.yuen@ntu.edu.sg).\\
\indent A. Al Hammadi is with the Technology Innovation Institute, 9639 Masdar City, Abu Dhabi, UAE (E-mail: ahmed.alhammadi@tii.ae).\\
\indent J. Yang is with the State Key Laboratory of Mobile Network and Mobile Multimedia Technology, Shenzhen, 518055, China. J. Yang is also with Wireless Product R\&D Institute, ZTE Corporation, Shenzhen 518057, China (E-mail: yang.jun10@zte.com.cn).\\
\indent M. Debbah is with Department of Electrical Engineering and Computer Science and the KU 6G Center, Khalifa University, Abu Dhabi 127788, UAE (E-mail: Merouane.Debbah@ku.ac.ae).
}\vspace{-1mm}}

\maketitle
\thispagestyle{empty}
\pagestyle{empty}

\begin{abstract}
\textcolor{black}{In integrated ground-air-space (IGAS) wireless networks, numerous services require sensing knowledge including location, angle, distance information, etc., which usually can be acquired during the beam training stage. On the other hand, IGAS networks employ large-scale antenna arrays to mitigate obstacle occlusion and path loss. However, large-scale arrays generate pencil-shaped beams, which necessitate a higher number of training beams to cover the desired space. These factors motivate our investigation into the IGAS beam training problem to achieve effective sensing services. To address the high complexity and low identification accuracy of existing beam training techniques, we propose an efficient hashing multi-arm beam (HMB) training scheme.} Specifically, we first construct an IGAS single-beam training codebook for the uniform planar arrays. Then, the hash functions are chosen independently to construct the multi-arm beam training codebooks for each AP. All APs traverse the predefined multi-arm beam training codeword simultaneously and the multi-AP superimposed signals at the user are recorded. Finally, the soft decision and voting methods are applied to obtain the correctly aligned beams only based on the signal powers. In addition, we logically prove that the traversal complexity is at the logarithmic level. Simulation results show that our proposed IGAS HMB training method can achieve 96.4\% identification accuracy of the exhaustive beam training method and greatly reduce the training overhead.
\end{abstract}
	
\begin{IEEEkeywords}
Beam training, integrated ground-air-space, sparsity, hashing, multi-arm beam, soft decision, voting mechanism.
\end{IEEEkeywords}
\vspace{-6mm}
\section{Introduction}
\vspace{1.5mm}
\textcolor{black}{As we enter the realm of 6G mobile communications, the landscape is rapidly shifting towards integrated ground-air-space (IGAS) networks\cite{8368236}. The shift to this integration is driven by the inherent advantages in terms of large coverage, high throughput, and strong resilience, one that can overcome the terrestrial limits of a traditional network. In such integrated networks, sensing services represent a critical constituent\cite{10556618}, which requires precise angle and distance information of the propagation channel. These are usually acquired during the beam training stage prior to the data transmission and offering other services\cite{Choi7786130,9410457}. Accurate and fast location identification is paramount across a myriad of applications including mobility management, packet routing, beam scheduling, and more. Thus, in the nascent stage of these 6G era network integrations, examining the importance and methodologies of improved sensing services has become an invaluable avenue of exploration.}
\par
\textcolor{black}{Typically, operational frequencies for ground-air-space communications are assigned at the millimeter-wave (mmWave) band, which offers higher communication frequencies than conventional frequency bands\cite{Jameel8594703}. However, short wavelengths make mmWave signals prone to being absorbed or blocked by obstacles such as buildings, trees, and even rain, resulting in high path loss and low communication coverage\cite{Niu07228,Wei7000981}. To mitigate the substantial losses and ensure reliable communication, a promising approach is forming highly directional beams using large-scale antenna arrays, to achieve high gain and spatial resolution.} %
\par
However, this means that the alignment of the transmitter and receiver beams becomes critical. Even small misalignments can result in signal degradation or loss. 
Besides, high spatial resolution results in a heavy beam training overhead, which is a major concern in mmWave systems. In the beam management procedure of the 802.11ad mmWave standard, access points (AP) typically sweep the entire angular space alternately to establish the aligned beam connections\cite{Jog2019ManytoManyBA}. The overhead associated with beam training is influenced by factors such as spatial resolution, the number of APs, and the dynamics of the propagation environment\cite{Va7742901}. To address these difficulties, we focus on beam training in IGAS networks and propose an efficient training method. In the following, we first review the relevant research works and then summarize our contributions.
\vspace{-3mm}
\subsection{Related works}
\vspace{0mm}
\textcolor{black}{\textit{Positioning in IGAS networks.} Location awareness is essential for the location-based services, and there have been many works investigating positioning in IGAS networks. In \cite{8917634}, an AI-based self-learning method was proposed for antenna pointing and mobile tracking. Utilizing unsupervised learning, high-precision position information can be acquired when users move across different environments. In \cite{6236174}, a cooperative positioning method based on time-division multiplexing was proposed, which provided vehicle position information. \cite{9399094} proposed a novel passive location parameter estimator using multiple satellites for the moving aerial target, which can effectively and precisely estimate the location parameters. In \cite{Wang1267608}, an unmanned aerial vehicle (UAV)-assisted target localization system was developed based on region division. \cite{8336153} presented a drone-based indoor people localization approach using ultrawideband signals. In \cite{8849936}, a UAV-aided ground user localization network using 3D time of arrival measurement was constructed by integrating global navigation satellite systems and ground cellular networks. \cite{9448650} proposed the UAV-aided non-parametric belief propagation localization where all users and UAVs are treated as uncertain agents due to their mobility. \cite{8422460} gave the Cram\'er–Rao lower bound of the ground user relative position errors in the UAV-aided localization network and developed a two-stage localization algorithm.}
\par
\textcolor{black}{\textit{Beam training in IGAS networks.} Accurate channel state information (CSI) is essential for achieving successful beam training. However, due to the variability of the wireless channel, obtaining precise CSI at all times, particularly in low signal-to-noise ratio (SNR) environments, is a challenge\cite{9264122}. In recent years, researchers have proposed a variety of beam training algorithms to align the directions of received and transmitted beams in the far field.}
\par
1) Priori information-assisted beam training\cite{Abdelreheem7847853,Choi7786130,Hashemi8101513,Alkhateeb6847111}: this algorithm utilized environmental priori information such as sensors, dedicated short-range communications (DSRC), and out-of-band information to configure mmWave links\cite{Choi7786130}, and the position information was considered to be a type of side information that can assist in establishing robust links in mmWave wireless communications. 
\par
2) Machine learning-based beam training\cite{Aviles7750625,pmlrv,10531779}: \cite{Aviles7750625} used a database containing past beam measurements at a specific location to determine the non-line of sight link beam formation direction. \cite{pmlrv} proposed a multi-armed bandit framework that can train multiple beam pairs in each attempt. In the multi-armed bandit setup, for arms with a high probability of obtaining a high reward, more attempts were required to obtain a certain cumulative reward.
\par
3) Blind beam training: The exhaustive training method, known as the most precise blind beam training technique, involved selecting the beam corresponding to the dominant path from a predefined codebook by exhaustively enumerating all possible beam combinations between the transmitter and receiver\cite{Junyi5262295}. However, this method necessitated traversing the entire beam space, resulting in significant delays and training overhead\cite{Wang9771330}. 
\par
In contrast, the hierarchical training method employed a multi-stage approach, dividing the beam space into two halves at each stage until the desired resolution was achieved\cite{Hur6600706}. While the hierarchical training method exhibited lower training overhead compared to the exhaustive training method, it had certain drawbacks. Utilizing a wide beam in the early stages can lead to a reduction in beamforming gain. Consequently, at low SNRs, the correct wide beam may not be accurately recognized, and the optimal beam direction may not be ultimately detected. Additionally, this method suffered from inherent error propagation problems\cite{Noh7417848}. Moreover, since comparative judgments of identification results were required at each stage to determine the training sub-codebooks for subsequent stages, additional delays were introduced as well.
\par
In addition to the single-beam training method, there are a few works investigating multi-arm beam training, which offered noticeable advantages in terms of complexity. \cite{You9129778} proposed the equal interval multi-arm beam (EIMB) training method, which utilized a predetermined sequence of multiple beams, and obtained the aligned beam direction through ensemble operations after multiple iterations. Agile-Link was presented in \cite{hassanieh2018fast}, which was a phased-array mmWave system that found the best beam alignment without scanning the entire space. In \cite{Xu2024LowComplexityBT}, the authors proposed a novel hashing multi-arm beam training scheme that reduces the training complexity to the logarithmic order with high accuracy.
\par
The aforementioned training techniques generally target the far-field region. However, the IGAS channels contain both angle and distance information, whereas conventional far-field beam training methods experience significant performance degradation. This is mainly due to the absence of angular domain sparsity in the IGAS communications, which is a key assumption for the design of far-field beam training codebooks. The proximity effect disrupts the sparsity of the far-field steering vectors, rendering the far-field codebook mismatched with the near-field channel. Consequently, it is necessary to exploit the beam training method by taking full consideration of near-field communication characteristics. \cite{Vives4244595} presented a modeling approach that utilized the near-field scanning method to measure the amplitude and phase of the magnetic field. \cite{Zhang9913211} proposed a novel two-phase beam training method, which decomposed the two-dimensional search into two sequential phases. \cite{Wei9810144} created a near-field codebook according to the characteristics of the near-field channel model. Additionally, a corresponding hierarchical near-field beam training scheme was proposed. The analysis in \cite{Cui9957130} indicated that despite the reduction in array gain, the near-field beam split effect can still contribute to achieving rapid near-field beam training. \cite{Liu10130629} proposed two deep learning-based near-field beam training schemes, and employed deep residual networks to determine the optimal near-field RIS codeword. \cite{10163797} trained the deep neural network by performing beam training based on the near-field codebook, where the optimal angle and distance were jointly predicted. \cite{you2023nearfield} provided an overview of near-field communications and introduced various applications of extremely large-scale arrays in both outdoor and indoor scenarios. Additionally, promising solutions were proposed to address these challenges.
\vspace{-4mm}
\subsection{Contributions}
\vspace{0.8mm}
\textcolor{black}{The main contribution of this paper is that we propose a low-complexity, high-accuracy hashing multi-arm beam (HMB) training scheme for IGAS networks. For the sake of illustration, we discuss beam training for the downlink in the near field, and it is worth noting that the beam training problem for the uplink can be addressed similarly. }
\par
\textcolor{black}{Specifically, we draw inspiration from the construction principle of far-field discrete Fourier transform (DFT) codebooks and construct an IGAS  single-beam training codebook that aims to minimize interference between the training beams. To further enhance performance through multi-arm beam techniques, we employ independent hash functions and jointly design the antenna responses. This enables us to construct the per-AP HMB codebook. Then, all APs traverse the predefined multi-arm beam codewords simultaneously and user's superimposed received signal power is recorded. Lastly, only the signal power is needed to achieve beam alignment by the soft decision and voting mechanism.}
\par
Furthermore, we analyze the training complexity of the proposed method and prove that the traversal complexity is at the logarithmic level. Simulation results show that our IGAS HMB training method achieves a notable enhancement in the identification accuracy of IGAS beam training. Specifically, the method achieves 96.4\% accuracy compared to the exhaustive beam training method while significantly reducing the training overhead. In addition, we also validate the applicability of our method under far-field conditions.
\par
The rest of the paper is organized as follows. Section II presents the channel and signal models that are relevant to the investigated IGAS networks. In Section III, we provide a detailed explanation of the IGAS single-beam codebook generation method and the process of forming the hashing-based multi-arm beam. Section IV presents the functioning of the decision and voting mechanisms and a theoretical proof of the complexity. Section V presents the simulation results, showcasing the performance of the proposed beam training technique. Finally, in Section VI, we conclude the paper.

\vspace{-2mm}
\section{System Model}
\begin{figure}[t]\vspace{2mm}
	\begin{center}
			\centerline{\includegraphics[width=0.45\textwidth]{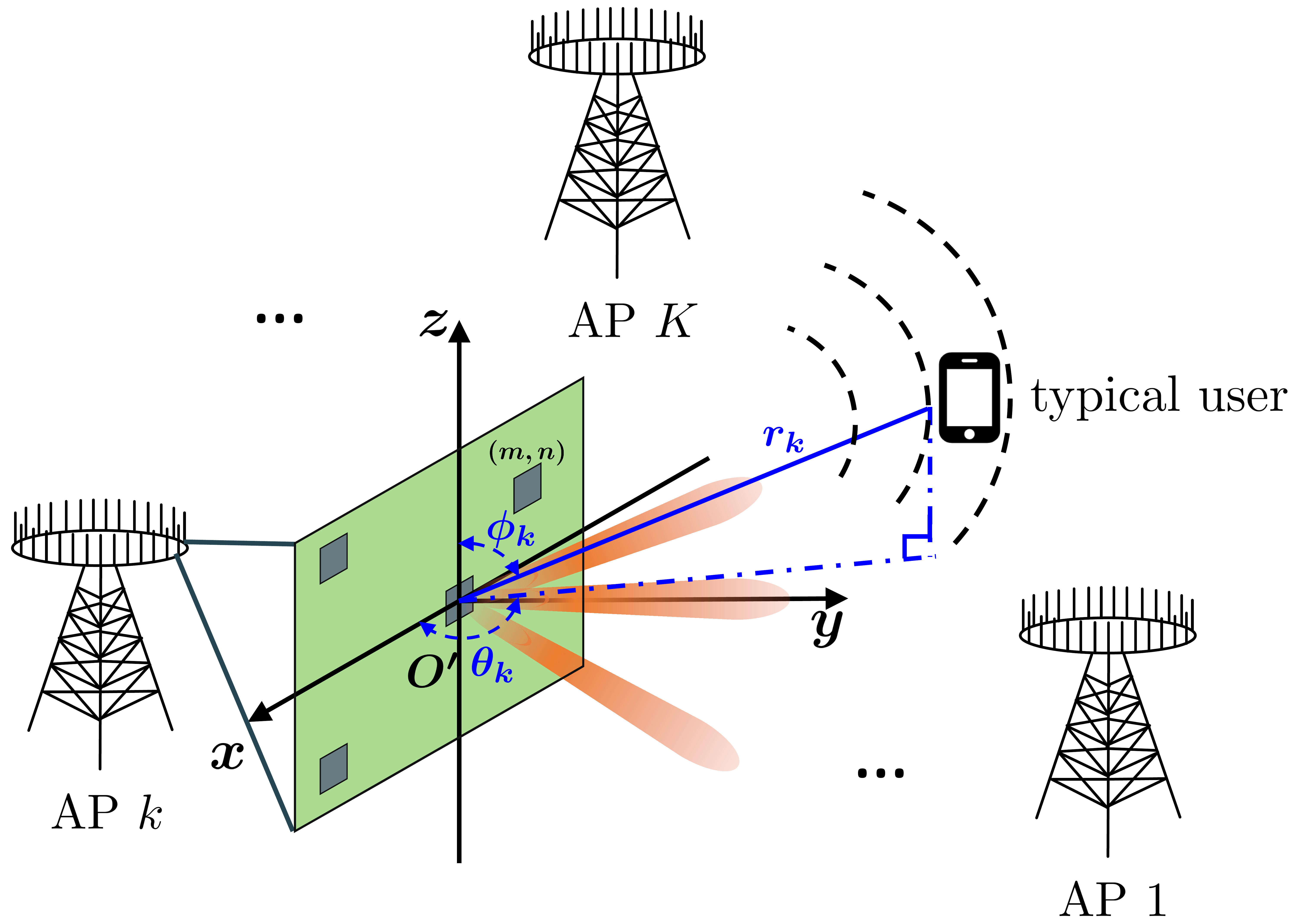}}  \vspace{-2mm}
			\caption{\textcolor{black}{Downlink IGAS mmWave scenario with $K$ APs and several users. }}
			\label{fig:scene} \vspace{-7mm}
		\end{center}
\end{figure}

\vspace{2mm}
\textcolor{black}{We consider a downlink IGAS mmWave scenario distributing $K$ APs and several users.} As shown in Fig.~\ref{fig:scene}, the antenna arrays of the APs are deployed in the $xz$-plane, each employing a hybrid precoding architecture that equips $V$ radio-frequency (RF) chains and a $M\times N$ antenna uniform planar array (UPA), where $V\ll MN$. Each user device is equipped with a single antenna. The central wavelength, horizontal antenna spacing, vertical antenna spacing, and operating frequency are $\lambda_c$, $d_x$, $d_z$, and $f_c$, respectively. The coordinate of the $(m,n)$-th antenna element of the $k$-th AP is $(x_{k,n},y_k, z_{k,m})$, where $x_{k,n}=r_k\cos\theta_k\sin\phi_k+nd_x $, $y_k=r_k\sin\theta_k\sin\phi_k$, $z_{k,m}=r_k\cos\phi_k+md_z $, $n=1-\frac{N+1}{2},...,N-\frac{N+1}{2}$, $m=1-\frac{M+1}{2},...,M-\frac{M+1}{2}$. The variables $r_k$, $\theta_k$ and $\phi_k$ denote the distance, azimuth angle, and elevation angle from the point $O'$ to the AP $k$, respectively.
\par
Let $\mathbf{h}_k\in \mathbb{C}^{MN\times 1}$ denote the channel from the AP $k$ to the user, the received signal $y$ at the user can be expressed as
\begin{equation}
	y=\sum\limits_{k=1}^K\mathbf{h}_k^H\mathbf{F}_{RF_k}\mathbf{f}_{BB_k}x+n_k,
\end{equation}
where $x$ denotes the transmit symbol with power $P_0$, $n_k\sim\mathcal{CN}(0,\sigma^2)$ denotes the Gaussian additive white noise, $\mathbf{F}_{RF_k}\in \mathbb{C}^{MN\times V}$ and $\mathbf{f}_{BB_k}\in \mathbb{C}^{V\times1}$ denote the analog beamformer and digital precoder at the $k$-th AP. 
\par
\textcolor{black}{In order to design the AP beamforming vector $\boldsymbol{a}\triangleq\mathbf{F}_{RF_k}\mathbf{f}_{BB_k}$ for training, we need to represent the channel $\mathbf{h}_k$ more concretely. We consider the near-field channel since it has the same form containing both angle and distance information. Different from the traditional plane-wave assumption for the far field, the propagating wavefront of the electromagnetic wave in the near field is approximated as a spherical surface. Moreover, the near field can be divided into the reflected region and the radiated region. Specifically, the former is the near-field region immediately adjacent to the antenna aperture. It is typically defined as $Z<0.62\sqrt{\frac{2D^2}{\lambda}}$, where $D$ represents the aperture size and $\lambda$ denotes the wavelength. In this region, no energy is externally radiated since the electric and magnetic fields convert between each other, resulting in energy storage and release. Usually, this region is relatively small, and the wave intensity exponentially decreases with distance. }
\par
The latter, which we focus on in this paper, is also named the Fresnel region. It is typically defined as $0.62\sqrt{\frac{2D^2}{\lambda}}<Z<\frac{2D^2}{\lambda}$. In this region, the electromagnetic wave is free from the constraints of the antenna and propagates into space. In addition, its cross-polarised electric field components result in the synthesis of an elliptically polarized wave in a plane parallel to the propagation direction. Thus, it is related to the angular distribution and distance information. In the following, we use the term "near field" for simplicity to refer to the radiated near field.
\par
Based on the spherical wave assumption, the near-field channel can be expressed as\vspace{-1mm}
\begin{equation}\label{eq:channel model1}
	\mathbf{h}_k=\sqrt{MN}\sum\limits_{l=1}^L\beta_l e^{-j\psi_l} \mathbf{g}_l,\vspace{-1mm}
\end{equation}
where $\beta_l$ and $\psi_l$ denote the complex path gain and phase shift of the $l$-th path. In the line of sight (LoS) link-dominated channel, $|\beta_1|\gg|\beta_l|, l\neq 1 $. As a result, we can approximate $\mathbf{h}_k$ as follows \vspace{-1mm}
\begin{equation}
    \mathbf{h}_k\approx\sqrt{MN}\beta_1 e^{-j\frac{2\pi}{\lambda_c}r_k}\mathbf{g}_1,
\end{equation}
where $\beta_1=\frac{\sqrt{\rho_0}}{r_k}$ and $\rho_0$ represents the reference channel power gain at a distance of $1$ meter. It can be observed that when $\boldsymbol{a}(\theta_s,\phi_s,r_s)=\mathbf{g}_1$, the received signal power from the $k$-th AP can be maximized. In other words, it is possible to simplify the beamforming problem to solve the angle
and distance information of the dominated path. 
\par
Thus, an intuitive idea is to design the training beam based on the structure of $\mathbf{g}_1$. Specifically, The vector $\mathbf{g}_1$ can be expressed as\par
\begin{align}\label{eq:g_1}
	&\mathbf{g}_1=\frac{1}{\sqrt{MN}} [e^{-j\frac{2\pi}{\lambda_c}(D(-\frac{M-1}{2},-\frac{N-1}{2})-r_k)},...,\nonumber\\\quad\quad&e^{-j\frac{2\pi}{\lambda_c}(D(m,n)-r_k)},...,e^{-j\frac{2\pi}{\lambda_c}(D(\frac{M-1}{2},\frac{N-1}{2})-r_k)}]^T,\\[2mm]
&D(m,n)=\left[(r_k\cos\theta_k\sin\phi_k+nd_x)^2\right.\nonumber\\\quad&\left.+(r_k\sin\theta_k\sin\phi_k)^2+(r_k\cos\phi_k+md_z)^2\right]^\frac{1}{2},
\end{align}
where $D(m,n)$ represents the distance between the $(m,n)$-th antenna of the $k$-th AP and the user.

 In the near-field region, assuming that $d_x\ll r_k$ and $d_z \ll r_k$, we can approximate the distance $D(m,n)$ using the second-order Taylor expansion\cite{Liu10220205}. The approximation is given by 
\begin{align}
    D(m,n)=&\left(r_k^2+n^2d_x^2+m^2d_z^2\right.\nonumber\\&\left.+2r_knd_x\cos\theta_k\sin\phi_k+2r_kmd_z\cos\phi_k\right)^\frac{1}{2}\nonumber\\
		\approx& r_k+nd_x\cos\theta_k\sin\phi_k+\frac{n^2d_x^2(1-\cos^2\theta_k\sin^2\phi_k)}{2r_k}\nonumber\\&+md_z\cos\phi_k+\frac{m^2d_z^2\sin^2\phi_k}{2r_k}.
\end{align}
\par
Thus, the phase of $\mathbf{g}_1$ can be decomposed into two parts that are only related to $m$ and $n$, respectively. These parts are given by $\frac{2\pi}{\lambda_c}(nd_x\cos\theta_k\sin\phi_k+\frac{n^2d_x^2(1-\cos^2\theta_k\sin^2\phi_k)}{2r_k})$ and $\frac{2\pi}{\lambda_c}(md_z\cos\phi_k+\frac{m^2d_z^2\sin^2\phi_k}{2r_k})$. Consequently, we obtain
\begin{subequations}\label{eq:channel model2}
	\begin{align}
    \mathbf{g}_1=\mathbf{v}_x(\theta,\phi,r)\otimes\mathbf{v}_z(\phi,r),
\end{align}
	\begin{equation}
		[\mathbf{v}_x(\theta,\phi,r)]_n=e^{-j\frac{2\pi}{\lambda_c}(nd_x\cos\theta_k\sin\phi_k+\frac{n^2d_x^2(1-\cos^2\theta_k\sin^2\phi_k)}{2r_k})},
	\end{equation}   
	\begin{equation}
		[\mathbf{v}_z(\phi,r)]_m=e^{-j\frac{2\pi}{\lambda_c}(md_z\cos\phi_k+\frac{m^2d_z^2\sin^2\phi_k}{2r_k})}.\vspace{-5mm}
	\end{equation} 
\end{subequations}

\vspace{2mm}
\section{\textcolor{black}{IGAS Codebook Generation}}
\vspace{2mm}
\textcolor{black}{In this section, we focus on designing training beams with minimal interference as possible. To address the mismatch between the existing far-field codebook and the IGAS beam training, we begin by designing a single-beam training codebook applicable to the IGAS network. Subsequently, we generate a multi-arm beam training codebook using the hashing method.}
\vspace{-2.8mm}
\subsection{\textcolor{black}{IGAS Single-Beam Codebook}}
\vspace{1mm}
In the far field, the scattering paths are limited and the far-field steering vectors are primarily angle-dependent, thus channel sparsity in the angle domain can be achieved by the DFT. By discretizing the angles and using the corresponding discrete Fourier vectors as orthogonal bases, the channel can be transformed into the angular domain. This transformation facilitates beam training by reducing the interference between training beams. \textcolor{black}{However, the IGAS channel model in (\ref{eq:channel model2}) reveals that the channel $\mathbf{h}_k$ is nonlinear concerning the antenna subscripts $m$ and $n$. In contrast to a discrete Fourier vector, $\mathbf{h}_k$ can be jointly described by several far-field Fourier vectors. Consequently, the energy of an IGAS path component is not concentrated at a single angle but leaks to multiple adjacent angles\cite{9685542}. This leakage indicates significant interference between angle-discretized training beams, resulting in substantial performance degradation when the DFT codebook is applied in the IGAS beam training.}
\par
Even though the IGAS channels are not sparse in the angular domain, there are finite paths that are compressible in the polar domain. To leverage this compressibility, we aim to design the beamforming vector $\boldsymbol{a}(\theta_s,\phi_s,r_s)\in\mathbb{C}^{1\times MN}$ as a set of orthogonal bases that exhibit the IGAS sparsity. This can be achieved by designing an angle- and distance-sampling approach. The general expression for $\boldsymbol{a}(\theta_s,\phi_s,r_s)$ is given by
\begin{align}\label{eq:dist_approx}
    \boldsymbol{a}(\theta_s,\phi_s,r_s)=&\frac{1}{\sqrt{MN}}\left[e^{j\frac{2\pi}{\lambda_c}(D^s(-\frac{M-1}{2},-\frac{N-1}{2})-r_s)},...,\right.\nonumber\\&\left.e^{j\frac{2\pi}{\lambda_c}(D^s(m,n)-r_s)},...,e^{j\frac{2\pi}{\lambda_c}(D^s(\frac{M-1}{2},\frac{N-1}{2})-r_s)}\right],\\
    D^s(m,n)=&\left[(nd_x-r_s\cos\theta_s\sin\phi_s)^2+(r_s\sin\theta_s\sin\phi_s)^2\right.\nonumber\\&\left.+(md_z-r_s\cos\phi_s)^2\right]^\frac{1}{2}
\end{align}
where $D^s(m,n)$ represents the distance from the $(m,n)$-th antenna of the AP to the sampling point $(\theta_s,\phi_s,r_s)$.
\par
The principle of distance- and angle-sampling method is to minimize $\eta$ as small as possible,
\par\vspace{-2mm}
\begin{equation}
	\eta\triangleq\max\limits_{p\neq q} f(\theta_p,\theta_q,\phi_p,\phi_q,r_{p},r_{q}),
\end{equation}
where $f(\theta_p,\theta_q,\phi_p,\phi_q,r_{p},r_{q})=|\boldsymbol{a}(\theta_p,\phi_p,r_p)\boldsymbol{a}(\theta_q,\phi_q,r_q)^H| $ denotes the projection between the two steering vectors at the sampled point $(\theta_p,\phi_p,r_p)$ and $(\theta_q,\phi_q,r_q)$. To obtain a closed-form expression, we make an approximation similar to (\ref{eq:dist_approx}) and obtain (\ref{eq:ff}) and (\ref{eq:f_m}) in the next page.
\begin{figure*}[ht]
    \centering\vspace{-5mm}
    \begin{subequations}\label{eq:ff}
\begin{align}\label{eq:f}
        f(\theta_p,\theta_q,\phi_p,\phi_q,r_{p},r_{q})=&|\frac{1}{MN}\sum\limits_{m=1-\frac{M+1}{2}}^{M-\frac{M+1}{2}}\sum\limits_{n=1-\frac{N+1}{2}}^{N-\frac{N+1}{2}} e^{j\frac{2\pi}{\lambda_c}[(D^p(m,n)-r_p)-(D^q(m,n)-r_q)]}|\nonumber\\
        =&|\frac{1}{MN}\sum\limits_{m=-\frac{M-1}{2}}^{\frac{M-1}{2}}\sum\limits_{n=-\frac{N-1}{2}}^{\frac{N-1}{2}} e^{j\frac{2\pi}{\lambda_c}(g_m(\phi_p,\phi_q,r_{p},r_{q})+g_n(\theta_p,\theta_q,\phi_p,\phi_q,r_{p},r_{q}))}|\nonumber\\
        =&|\frac{1}{N}\sum\limits_{n=-\frac{N-1}{2}}^{\frac{N-1}{2}}e^{j\frac{2\pi}{\lambda_c}g_n(\theta_p,\theta_q,\phi_p,\phi_q,r_{p},r_{q})}\frac{1}{M}\sum\limits_{m=-\frac{M-1}{2}}^{\frac{M-1}{2}}e^{j\frac{2\pi}{\lambda_c}g_m(\phi_p,\phi_q,r_{p},r_{q})}|\textcolor{black}{,}
    \end{align}
\begin{equation}\label{eq:g_m}
        g_m(\phi_p,\phi_q,r_{p},r_{q})=-md_z(\cos\phi_p-\cos\phi_q)+\frac{m^2d_z^2\sin^2\phi_p}{2r_p}-\frac{m^2d_z^2\sin^2\phi_q}{2r_q}\textcolor{black}{,}
    \end{equation}
    \begin{align}\label{eq:g_n}
        g_n(\theta_p,\theta_q,\phi_p,\phi_q,r_{p},r_{q})=&-nd_x(\cos\theta_p\sin\phi_p-\cos\theta_q\sin\phi_q)\nonumber\\
        &+\frac{n^2d_x^2(1-\cos^2\theta_p\sin^2\phi_p)}{2r_p}-\frac{n^2d_x^2(1-\cos^2\theta_q\sin^2\phi_q)}{2r_q}\textcolor{black}{,}
    \end{align}	
\end{subequations}
\begin{equation}\label{eq:f_m}
    f_m\triangleq\frac{1}{M}\sum\limits_{m=-\frac{M-1}{2}}^{\frac{M-1}{2}}e^{j\frac{2\pi}{\lambda_c}g_m(\phi_p,\phi_q,r_{p},r_{q})}\textcolor{black}{.}\vspace{-3mm}
\end{equation}
{\noindent} \rule[-10pt]{17.5cm}{0.05em}\\
\end{figure*}
\par
We observe that (\ref{eq:g_m}) (and similarly in (\ref{eq:g_n})) is only related to the index $m$ ($n$). Besides, when $f_m=0$, the function $f(\theta_p,\theta_q,\phi_p,\phi_q,r_{p},r_{q})$ evaluates to zero. Therefore, the goal is to make $f_m$ converge to zero as closely as possible. To achieve this, we can decouple (\ref{eq:g_m}) into two parts: one part containing only angle information, denoted as $g_m^a\triangleq-md_z(\cos\phi_p-\cos\phi_q)$, and another part containing both angle and distance information, denoted as $g_m^b\triangleq\frac{m^2d_z^2\sin^2\phi_p}{2r_p}-\frac{m^2d_z^2\sin^2\phi_q}{2r_q}$. In the subsequent analysis, we isolate the angular and distance components. 
\par 
Firstly, considering the case when $g_m^b=0$, we design the angular sampling method based on $g_m^a$, which implies that $\frac{\sin^2\phi}{2r}$ is a constant. In this scenario, the term $f_m$ only depends on $g_m^a$, which can be expressed as
\begin{equation}
    \begin{split}\label{eq:part 1}
        f_m=\frac{1}{M}\sum\limits_{m=-\frac{M-1}{2}}^{\frac{M-1}{2}} e^{j\frac{2\pi md_z}{\lambda_c}(-md_z(\cos\phi_p-\cos\phi_q))}\\
        =|\frac{\sin(\frac{\pi Md_z}{\lambda_c}(\cos\phi_q-\cos\phi_p))}{M\sin(\frac{\pi d_z}{\lambda_c}(\cos\phi_q-\cos\phi_p))}|.
    \end{split}
\end{equation}
\par
It can be found that (\ref{eq:part 1}) is independent of the distances $r_p$ and $r_q$, which is exactly equivalent to the projection between two far-field steering vectors. The zero points of (\ref{eq:part 1}) satisfy the equation $\cos\phi_q-\cos\phi_p=\frac{2i}{M}$, where $i=1,..,M-1$. Based on this observation, we can sample the angle $\cos\phi_s$ at equal intervals of $\frac{2}{M}$. Specifically, we can set $\cos\phi_s=\frac{2s-M-1}{M}$, where $s=1,...,M$. This angular sampling method ensures that the angles cover the range of possible values for the aligned angle representations.
\par
Afterward, we obtain the distance sampling method based on $g_m^b$ to ensure accurate recovery of the transmitted information $x$. This is a fundamental difference between the angular domain representation and the polar domain representation. Specifically, the projection $f(\theta_p,\theta_q,\phi_p,\phi_q,r_{p},r_{q})$ can be expressed as \vspace{-2mm}
\begin{equation}
\begin{split}
    &f(\theta_p,\theta_q,\phi_p,\phi_q,r_{p},r_{q})\\
    =&|\frac{1}{M}\sum\limits_{m=-\frac{M-1}{2}}^{\frac{M-1}{2}} e^{j\frac{2\pi}{\lambda_c}
    m^2d_z^2(\frac{\sin^2\phi_p}{2r_p}-\frac{\sin^2\phi_q}{2r_q})}|\\
    \approx& |\frac{1}{M}\int_{-\frac{M}{2}}^{\frac{M}{2}} e^{j\frac{2\pi}{\lambda_c}
    m^2d_z^2(\frac{\sin^2\phi_p}{2r_p}-\frac{\sin^2\phi_q}{2r_q})}\mathrm{d} m|.
\end{split}
\end{equation}
\par
By substituting
\begin{subequations}
    \begin{equation}
        t=\sqrt{\frac{2(\frac{\sin^2\phi_p}{2r_p}-\frac{\sin^2\phi_q}{2r_q})}{\lambda_c}}md_z,
    \end{equation}
    \begin{equation}
    \zeta=\sqrt{\frac{(\frac{\sin^2\phi_p}{2r_p}-\frac{\sin^2\phi_q}{2r_q})}{\lambda_c}}Md_z,
    \end{equation}
\end{subequations}
we have
\begin{equation}
\begin{split}
    &f(\theta_p,\theta_q,\phi_p,\phi_q,r_{p},r_{q})\\
    =&|\int_0^{\zeta} \frac{1}{\zeta} e^{j\frac{\pi}{2}t^2}\mathrm{d}t|\\
    \overset{(a)}{=}&|\frac{C(\zeta)+jS(\zeta)}{\zeta}|,
\end{split}
\end{equation}
where the equal sign $(a)$ is due to the fact that $C(x)=\int_0^x \cos(\frac{\pi}{2}t^2)dt$ and $S(x)=\int_0^x \sin(\frac{\pi}{2}t^2)dt $ are Fresnel integrals.
\par 
Since the projection $f(\theta_p,\theta_q,\phi_p,\phi_q,r_{p},r_{q})$ decreases with increasing $\zeta$ and approaches 0, we set a threshold $\Delta$, and calculate the value of $\zeta_\Delta$ that ensures the projection is less than $\Delta$. Therefore,
\begin{equation}\label{eq:dist_sample}
|\frac{\sin^2\phi_p}{r_p}-\frac{\sin^2\phi_q}{r_q}|\geq\frac{2\lambda_c\zeta_\Delta^2}{M^2d_z^2}.
\end{equation}
\par
If we use the angular sampling method described earlier, and sample the distances according to (\ref{eq:dist_sample}), then the projection of the two steering vectors will be no larger than $\Delta$. Consequently, we can construct the observation vector $\mathbf{a}$ as orthogonal bases $\boldsymbol{a}(\theta_s,\phi_s,r_s)$. By doing so, we can extract the signal in the IGAS position $(\theta_s,\phi_s,r_s)$. To accurately capture the IGAS signal, we use the collection of $\boldsymbol{a}(\theta_s,\phi_s,r_s)$ based on all the sampling points to form the IGAS single-beam codebook $\mathbf{C} $, which can be expressed as\vspace{-2mm}
\begin{equation}
\mathbf{C}=[\boldsymbol{a}(\theta_1,\phi_1,r_1);...;\boldsymbol{a}(\theta_S,\phi_S,r_S)].
\end{equation}
\par
It is worth noting that the codebook construction method described above can also be applied to the far-field region when the distance sampling $r_s$ tends to infinity. It can be seen in the simulation results that the codebook constructed using this method has comparable performance to that of the DFT codebook.
\vspace{-3mm}
\subsection{Hashing Multi-Arm Beam Codebook Generation}
\vspace{1.5mm}
To reduce the beam training overhead while maintaining the identification accuracy, we consider the generation of multi-arm beams for training. The multi-arm beams point in multiple directions simultaneously and can be constructed by selecting codewords from the single-beam codebook $\mathbf{C}$ and combining them. 
The combination can be accomplished by hashing in computer science, by arranging the codewords from the single-beam codebook. Specifically, hashing is a technique commonly employed to store data based on its keyword or other attributes\cite{Wang7915742}. Here's an outline of the main ideas:
\par
1. Define the universe of keys: We denote the universe of keys as the collection of all codeword indexes $\mathcal{U}=\{0,1,...,N_C-1\}$, where $N_C$ represents the number of codewords in the single-beam codebook.
\par
2. Define the hash values: We define the interpreted hash values as $\mathcal{T}=\{0,1,...,B-1\}$, where a hash function $h:\mathcal{U}\to\mathcal{T}$ maps each key (codeword index) to the interpreted hash value within the desired range.
\par
3. Randomly select hash functions: From a family $\mathcal{H}=\{h_1,h_2,...,h_{|\mathcal{H}|}\}$ of hash functions, we randomly select a subset of hash functions to be used in the combination. The family $\mathcal{H}$ consists of distinct hash functions, and $|\mathcal{H}|$ represents the total number of distinct functions in the family. The construction of the family $\mathcal{H}$ is based on the following theorem.
\begin{theorem}\label{theorem:k-wise}
For the Galois field $GF(N_C)$, we construct a $k$-wise independent map from $\mathcal{U}$ to $\mathcal{T}$ as follows:
\par
Pick $k$ random numbers $a_0,a_1,...,a_{k-1}\in GF(N_C)$. For any $x\in\mathcal{U}$, 
\begin{equation}
    h(x) \triangleq a_0 + a_1x + a_2x^2 + ... + a_{k-1}x^{k-1}\mod\ {B},
\end{equation}
where the calculations are done over the field $GF(N_C)$, and $|\mathcal{H}|=N_C^{k}-N_C$, because $a_1,a_2,... . a_{k-1}$ cannot equal to zero at the same time.
\end{theorem}
\par
\textcolor{black}{\textit{Proof:} See Appendix A.}
\vspace{1mm}
\par
Next, we can utilize the hashing results to generate the multi-arm beams. This process involves randomly selecting a hash function from the family $\mathcal{H}$ and computing the hash values of all elements $x\in\mathcal{U}$ which serve as storage addresses. Consequently, each bucket $\mathbf{d}_b$ comprising $R=N_C/B$ keywords represents a multi-arm beam. Thus, we can represent the multi-arm beams generated by hashing as follows:
\begin{equation}
	\mathbf{D}=[\mathbf{d}_1;...;\mathbf{d}_B],\quad \mathbf{d}_b=[{d}_b^1,...,{d}_b^R],
\end{equation}
where commas ($,$) and semicolons ($;$) denote the row separators and column separators, respectively.
\par
\begin{figure}[t]
	\centering  
	\includegraphics[width=1\linewidth]{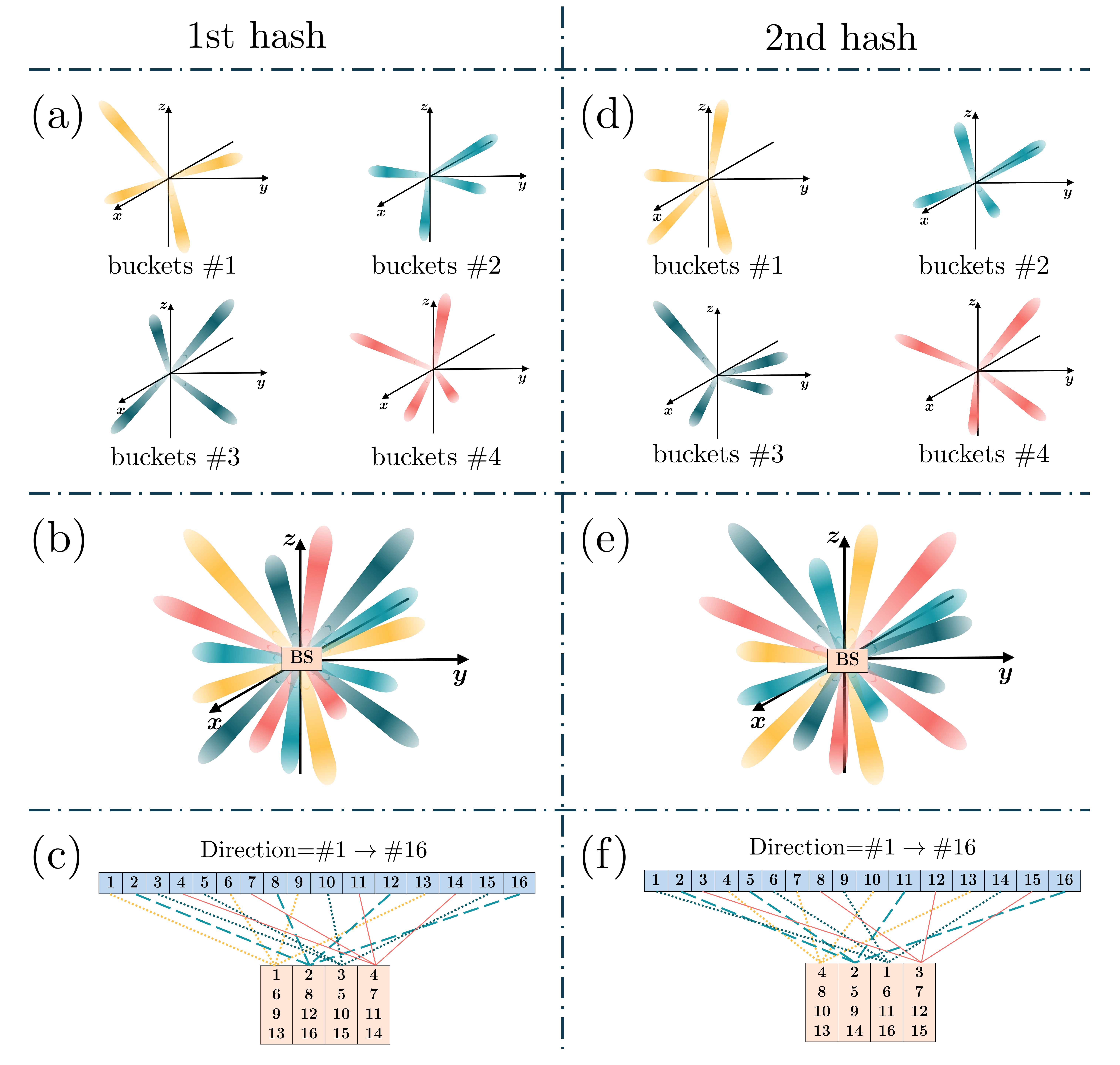}\vspace{-1mm}
	\caption{The schematic diagram for hashing implementation.}\label{fig:hash}\vspace{-2mm}
\end{figure}

Fig.~\ref{fig:hash} illustrates two times hash that hashing the codewords representing $16$ directions uniformly into $4$ multi-arm beams. Each multi-arm beam encompasses a total of $N_C/B = 4$ distinct directions. 
\par 
In the following, we design analog beamformer $\mathbf{F}_{RF_k}$ and digital precoder $\mathbf{f}_{BB_k}$ to form multi-arm beams. In contrast to the simple antenna partitioning approach\cite{You9129778}, we jointly design the response of all antennas to generate the multi-arm beam codebook $\tilde{\mathbf{C}}$ for training. Specifically, for each $\mathbf{d}_b$, we optimize the digital precoder $\mathbf{f}_{BB_k}$ to map the data stream to the designated RF chain, while the analog beamformer $\mathbf{F}_{RF_k}$ selects $V=R$ codewords from the single-beam codebook $\mathbf{C}$ based on the contained codeword indexes. This selection determines the beams to be transmitted through the designated RF chain. In other words, the analog beamformer controls the directionality and beamforming characteristics of the transmitted signal,
\vspace{-0mm}
\begin{equation}
	\mathbf{f}^b_{BB_k}(i,1)=\frac{e^{j\boldsymbol{\vartheta}(i)}}{\sqrt{V}},i=1,...,V,
\end{equation}
\begin{equation}
	\mathbf{F}^b_{RF_k}(:,i)=\mathbf{C}(\mathbf{d}_b^i,:)^T,i=1,...,V.
\end{equation}
\par
Therefore, the received power at the user from AP $k$ can be derived as (\ref{eq:y_K}). The normalized multi-arm beam radiation pattern of $\mathbf{d}_b$ is denoted by (\ref{eq:W}), while (\ref{eq:W'}) represents the normalized radiation pattern of the $s$-th single beam.
\begin{figure*}[t]
    \centering
    \begin{equation}\label{eq:y_K}
	\begin{split}
		|y_k|^2(\mathbf{d}_b,\boldsymbol{\vartheta})&\overset{(a)}{\approx}|\sqrt{P_0}|\beta_1|\sqrt{MN}e^{-j\frac{2\pi}{\lambda_c}r_k}\sqrt{W(r_k,\theta_k,\phi_k,\boldsymbol{\vartheta},\mathbf{d}_b)}  +n_k|^2
	\end{split}
    \end{equation}
    \begin{equation}\label{eq:W}
	W(r_k,\theta_k,\phi_k,\boldsymbol{\vartheta},\mathbf{d}_b)=\sum\limits_{i=1}^{V}\frac{1}{V} e^{j\boldsymbol{\vartheta}(i)}\sum\limits_{n=-\frac{N-1}{2}}^{\frac{N-1}{2}}\sum\limits_{m=-\frac{M-1}{2}}^{\frac{N-1}{2}} \frac{1}{MN} e^{-j\frac{2\pi}{\lambda_c}(D^{\mathbf{d}_b^i}(m,n)+D(m,n)-r_{\mathbf{d}_b^i}-r_k)}
    \end{equation}
    \begin{equation}\label{eq:W'}
	W'(r_k,\theta_k,\phi_k,s)=\sum\limits_{n=-\frac{N-1}{2}}^{\frac{N-1}{2}}\sum\limits_{m=-\frac{M-1}{2}}^{\frac{N-1}{2}} \frac{1}{MN} e^{-j\frac{2\pi}{\lambda_c}(D^{
			s}(m,n)+D(m,n)-r_{\mathbf{d}_b^i}-r_k)}
    \end{equation}
\end{figure*}
\par
Because different IGAS single beams interfere with each other in their respective main lobes, the $s$-th single beam has a main lobe region denoted by $\cos\phi'\in[\cos\phi_s-\frac{1}{M},\cos\phi_s+\frac{1}{M}]$, $\cos\theta'\in[\cos\theta_s-\frac{1}{N},\cos\theta_s+\frac{1}{N}]$, 
\begin{equation}
    r'\in[\frac{1+\frac{r_s\lambda_c\zeta_\Delta^2}{\sin^2\phi_sM^2d_z^2}}{\frac{1}{r_s}+\frac{2\lambda_c\zeta_\Delta^2}{\sin^2\phi_sM^2d_z^2}},\frac{1-\frac{r_s\lambda_c\zeta_\Delta^2}{\sin^2\phi_sM^2d_z^2}}{\frac{1}{r_s}-\frac{2\lambda_c\zeta_\Delta^2}{\sin^2\phi_sM^2d_z^2}}].
\end{equation}
\begin{figure*}[ht]\vspace{-5mm}
    \centering
    \begin{equation}\label{eq:delta_W}
	\delta_W(\boldsymbol{\vartheta},\mathbf{d}_b^i)\triangleq \int_{r'}\int_{\theta'}\int_{\phi'}|\frac{W(r_k,\theta_k,\phi_k,\boldsymbol{\vartheta},\mathbf{d}_b)-W'(r_k,\theta_k,\phi_k,\mathbf{d}_b^i)}{W'(r_k,\theta_k,\phi_k,\mathbf{d}_b^i)}|\mathrm{d} \phi\mathrm{d} \theta\mathrm{d}r\vspace{-2mm}
\end{equation}
{\noindent} \rule[-10pt]{17.5cm}{0.05em}\\
\end{figure*}
We define the unit deviation of $W(r_k,\theta_k,\phi_k,\boldsymbol{\vartheta},\mathbf{d}_b)$ and $W'(r_k,\theta_k,\phi_k,\mathbf{d}_b^i)$ as (\ref{eq:delta_W}).
Further, we define the average deviation between the multi-arm beam and single-beam radiation patterns within their respective main lobes as
\begin{equation}
	\delta_W(\boldsymbol{\vartheta},\mathbf{d}_b)\triangleq \frac{1}{V}\sum\limits_{i=1}^{V} \delta(\boldsymbol{\vartheta},\mathbf{d}_b^i).
\end{equation}
\par
Note that when the average deviation $\delta_W(\boldsymbol{\vartheta},\mathbf{d}_b)$ is sufficiently small, the main lobe of the sub-beam pattern closely resembles the corresponding single beam. In other words, the main lobe characteristics of the single beam are preserved within the multi-arm beam structure. Therefore, we can employ the aforementioned HMB combination $\mathbf{d}_b$ to minimize $\delta_W(\boldsymbol{\vartheta},\mathbf{d}_b)$ by adjusting $\boldsymbol{\vartheta}=\boldsymbol{\vartheta}_b$ appropriately. This adjustment allows us to generate a well-shaped multi-arm beam, constituting the $b$-th row of the Hashing multi-arm beam Codebook $\tilde{\mathbf{C}}(b,:)=\mathbf{F}^b_{RF_k}\mathbf{f}^b_{BB_k}$.
\vspace{-0mm}
\section{integrated ground-air-space Beam Traning}
\vspace{3mm}
The training process consists of two phases: the scanning phase and the beam identification phase. In the scanning phase, APs send training symbols using predefined multi-arm beams, each with the same power, until all the predefined multi-arm beams are traversed. In the beam identification phase, we employ a demultiplexing algorithm based on the soft decision to separate the superimposed signals from multiple APs. Afterward, we utilize the voting mechanism to obtain aligned beams for each AP.
\par
For a single AP, the soft decision and voting mechanism work as follows. Referring to the first hash illustrated in Fig.~\ref{fig:hash} (c), if a signal arrives from direction $9$, only bucket $1$ detects signal energy, while buckets $2$, $3$, and $4$ contain only the noise. This narrows the search space to the directions included in the first multi-arm beam. That is, the potential directions from which the signal originated are $1$, $6$, $9$, and $13$. Afterward, a second hashing step is performed, as shown in Fig.~\ref{fig:hash} (e). As the coefficients of the second hash function are randomized, it is unlikely for the hash results to match those of the previous hash function. In the second hash, only the second bucket exhibits high energy. This indicates that the signal arrives from one of the directions mapped to the third bucket, namely directions $2$, $5$, $9$, and $14$. Since the common candidate direction in both the first and second hashes is direction $9$, we obtain the $9$-th codeword as the aligned beam.
\par
In the case of multiple APs, to ensure accurate beam training, we employ a total of $L$ rounds of hash. Specifically, AP $k$ randomly selects $L$ distinct hash functions $h_1^k, h_2^k, \ldots, h_L^k$ from a predefined family $\mathcal{H}$ to obtain $\mathbf{D}_1^k, \mathbf{D}_2^k, \ldots, \mathbf{D}_L^k$. During the scanning phase, AP $k$ transmits training symbols using the codeword from its multi-arm beam codebooks $\tilde{\mathbf{C}}_1^k, \tilde{\mathbf{C}}_2^k, \ldots, \tilde{\mathbf{C}}_L^k$. As a result, a total of $Q = BL$ time slots are required for the scanning process, significantly reducing the complexity compared to the traditional method of scanning all APs. It yields $Q$ received signal power for the typical user, denoted as $\mathbf{P}=[P(1,1),...,P(l,b),...,P(L,B)]$. Here, the measurement $P(l,b)$ represents the power of the signal received by the $b$-th multi-arm beam during the $l$-th round of hash, corresponding to the $q = (l-1)B + b$-th time slot.\vspace{-1mm}
\begin{equation}
	 P(l,b)=|\sum\limits_{k=1}^K\mathbf{h}_k^H\mathbf{F}_{RF_k}\mathbf{f}_{BB_k}x+n_k|^2
\end{equation}
\par
In the following, we introduce the mechanism of the beam identification phase for the multi-AP case.
\vspace{-1mm}
\subsection{Beam Identification Phase}
\vspace{4mm}
Suppose the direction of the user to AP $k$ is denoted as  $\gamma_k\in\mathcal{U}$. According to (\ref{eq:k-wise}), the probability of two arbitrary APs observing this user simultaneously, i.e., $h_i(\gamma_i) = h_j(\gamma_j)$ for $i \neq j$, is given by $Pr(h_i(\gamma_i) = h_j(\gamma_j)) = \frac{1}{B^2}$. This probability is small enough such that the received signal of each time slot typically contains the signal from at most one AP. Furthermore, due to varying distances between different APs and the user, distinct channel gains are obtained. With the same transmit power, it results in distinguishable received signal powers, denoted as $\dot{P}_{m_1}>\dot{P}_{m_2}>...>\dot{P}_{m_K}$, where $m_k$ corresponds to the AP with the $k$-th strongest channel gain. Thus, the demultiplexing algorithm can be designed in conjunction with the soft decision. Specifically, we assign the $L$ time slots with the $(k-1)L\!+\!1$-th $\sim$ $kL$-th largest value in the received signal power vector $\mathbf{P}$ to AP $m_k$, which means\vspace{-1mm}
\begin{equation}\vspace{-2mm}
	\begin{split}
		\mathbf{q}_{m_k}=&\arg\max\limits_{(k-1)L+1:kL} descend(\mathbf{P})
	\end{split},
\end{equation} 
where $descend(\cdot)$ represents the operation that sorts the vector in descending order. The reason for taking $L$ numbers is that each AP sees the user in only one time slot per round. 
\par
Now that we distinguish the received signal power from different APs, we can proceed with a vote on $\tilde{\mathbf{D}}^k(\mathbf{q}_{m_k},:)$ to determine $\gamma_k$, where colon ($:$) denotes all the elements of the row/column. In this process, the voting can lead to a unique direction due to the universality property of the hash function family, which implies that for $\forall x\in\mathcal{U}$ and $\forall \alpha\in\mathcal{T}$, the probability $Pr[h(x) = \alpha]$ is equal to $\frac{1}{B}$. Equivalently, for $\forall x_1,x_2\in\mathcal{U}$, we have $Pr[h(x_1) = h(x_2)] = \frac{1}{B}$. Thus, given $x_1$ and $x_2$ that satisfy $h_i(x_1)=h_i(x_2)$, we have $Pr[h_{i'}(x_1) = h_{i'}(x_2)] = \frac{1}{B}$, where ${i'\neq i}$, $h_i,h_{i'}\in\mathcal{H}$. Moreover, due to the random selection of polynomial coefficients and the distinctness of functions within the hash function family, we can establish (\ref{eq:BL}).
\begin{figure*}[ht]
    \centering
    \begin{equation}\label{eq:BL}
	Pr(h_1(x_1)=h_1(x_2)\ \wedge\ h_2(x_1)=h_2(x_2)\ \wedge\ ...\ \wedge\ h_L(x_1)=h_L(x_2))=\frac{1}{B^L}\vspace{-5mm}
    \end{equation}
    {\noindent} \rule[-10pt]{17.5cm}{0.05em}
\end{figure*}
\par
As a result, the probability of two keywords being hashed to the same address simultaneously by multiple hash functions is sufficiently small. It ensures that multiple rounds of hash make the directions dispersed from each other. In other words, based on the randomness of the hash functions, when voting is conducted on arbitrary $L$ beams of the $k$-th AP, the resulting votes are scattered, and the direction receiving the highest number of votes approximately follows a uniform distribution. However, considering the demultiplexed time slot $\mathbf{q}_{m_k}$ containing the aligned beam, it is highly probable that the direction $\gamma_{m_k}$ will receive the highest number of votes. This is attributed to the fact that the aligned beam corresponds to the strongest received signal power, and the voting process is designed to identify the dominant direction based on the signal power.
\par
The detailed steps of the beam training are discussed in Algorithm \ref{alg:Framwork}. Firstly, in the scanning phase, all APs simultaneously send training symbols utilizing the predefined multi-arm beams with the same power $P_0$, until all the predefined multi-arm beams are traversed. Concurrently, all users listen to the channel using a quasi-omnidirectional beam. In the beam identification phase, we demultiplex the received multi-AP superimposed signal power $\mathbf{P}_u$ at user $u$ using a soft decision approach. This involves assigning the $L$ time slot, determined by the $(k-1)L+1$-th to $kL$-th largest values in $\mathbf{P}_u$, to the corresponding APs based on their indices, denoted as $m_k$. Subsequently, we perform a voting process on $\tilde{\mathbf{D}}^k(\mathbf{q}_{m_k},:)$ to determine the direction $\gamma^u_k$ with the highest number of votes. Finally, the aligned beam of AP $k$ for user $u$ is selected as the $\gamma^u_k$-th codeword in the single beam codebook $\mathbf{C}$, represented as $\mathbf{C}(\gamma_k^u,:)$.

\vspace{-2mm}
\subsection{Complexity Analysis}\label{sec:complexity}
\vspace{2mm}
\begin{theorem}\label{theorem:complexity}
Given the number of rounds of hash is $L=O(\mathrm{log}M_s)$, it is guaranteed that the probability of identification error is less than $\frac{1}{M_s}$.
\end{theorem}
\par
\textcolor{black}{\textit{Proof:} See Appendix B.}
\begin{algorithm}[t]\vspace{1mm}
	\caption{HMB Training}
	\label{alg:Framwork}
	\begin{algorithmic}[1]
		\Require 
		\Statex the multi-arm beams for all APs $\{\mathbf{D}^k_1,...,\mathbf{D}^k_L\}_{k=1}^K$
            \Statex the codebooks $\{\tilde{\mathbf{C}}_1^k, \tilde{\mathbf{C}}_2^k, \ldots, \tilde{\mathbf{C}}_L^k\}_{k=1}^K$
		\Statex the transmit signal $x$
		\Statex the number of APs $K$
		\Statex the number of hash rounds $L$
            \Statex the number of time slots $Q$
            \vspace{1mm}
		\Ensure 
		\Statex the aligned beam indexes $\{\bm{\gamma}^u\}_{u=1}^U$, $\bm{\gamma}^u=[\gamma^u_1,...,\gamma^u_K]$
            \Statex the aligned beams of APs corresponding to users 
		
		\vspace{2mm}
		\For{$q$ = 1 to $Q$}
		\State $\forall$ AP $k$ transmit $x$ by the $q$-th multi-arm beam in 
  $\{\tilde{\mathbf{C}}_1^k, \tilde{\mathbf{C}}_2^k, \ldots, \tilde{\mathbf{C}}_L^k\}$
		\State all users record the multi-AP superimposed received signal powers $\{\mathbf{P}_u\}_{u=1}^U$
		\EndFor
  \vspace{1mm}
		\For{($\forall$ user $u$) $k$ = 1 to $K$}
		\State $\mathbf{q}_{m_k}=\arg\max\limits_{(k-1)L+1:kL} descend(\mathbf{P}_u)$
		\State $\gamma^u_k\leftarrow$ most votes on $\tilde{\mathbf{D}}^k(\mathbf{q}_{m_k},:)$
		\State aligned beam of AP $k$ to user $u$ is $\mathbf{C}(\gamma_k^u,:)$
		\EndFor
	\end{algorithmic}
\end{algorithm}
\par
Thus, the number of rounds of hash is only related to the desired identification accuracy, a logarithmic level of rounds of hash is sufficient. By using a logarithmic number of rounds, the overhead of HMB training can be significantly reduced compared to the traditional exhaustive beam training.
\par
\textcolor{black}{Here we discuss the advantages of the HMB training compared with existing IGAS beam training methods. In contrast to single-beam training, multi-arm beam training can scan multiple directions simultaneously, greatly reducing the overhead of beams to cover the space. Both the hashing and the equal interval approaches use multi-arm beams. However, the EIMB method uses a predetermined multi-arm beam combination approach as well as a hard decision of threshold comparison, which are the main aspects that differentiate it from our method. The HMB training employs randomly selected hash functions, which can be perceived as introducing a perturbation to the deterministic combination approach, thereby rendering the combination random. This randomness ensures that the sub-beams are sufficiently dispersed, thereby reducing the impact of leakage interference between sub-beams, particularly after multiple rounds of voting.}
\par
\textcolor{black}{As illustrated in Fig.~\ref{fig:hash}, if there is interference from beam $1$ to beam $9$ in the first round, there will be no further interference from beam $1$ in subsequent rounds because $1$ and $9$ are no longer in the same multi-arm beam. By employing an $L$-round voting mechanism, each round carries equal weight, thus mitigating the influence of errors from individual rounds when $L$ is relatively large. Moreover, since $L$ is complexity dependent, we indicate that $L$ is on the order of $O(\mathrm{log} M_s)$ when the identification error is approximately $1/M_s$. That is, we sacrifice some accuracy compared to the exhaustive approach but greatly reduce the training overhead to $B\mathrm{log}MN$.}
\par
\textcolor{black}{Another distinction lies in the soft decision. We consider the relative value, unlike the EIMB method which considers the absolute value. This approach makes it less susceptible to noise and variations in practical power strength. Hence, our method can achieve higher accuracy and greater robustness.}

\begin{figure}[t]\vspace{-3mm}
	\begin{center}
			\centerline{\includegraphics[height= 0.382\textwidth]{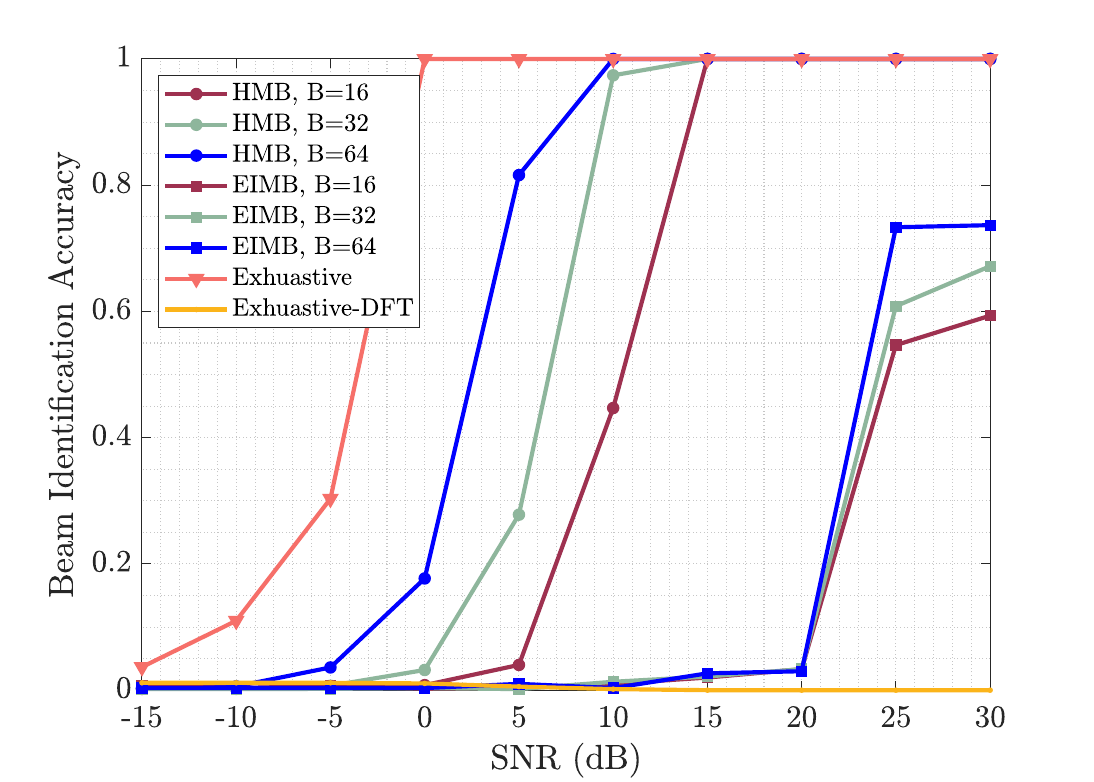}}  \vspace{-1mm}
			\caption{Success beam identification accuracy versus the SNR.  }
			\label{fig:accuray} \vspace{-6mm}
		\end{center}
\end{figure}
\begin{figure}[t]
	\begin{center}
			\centerline{\includegraphics[height= 0.382\textwidth]{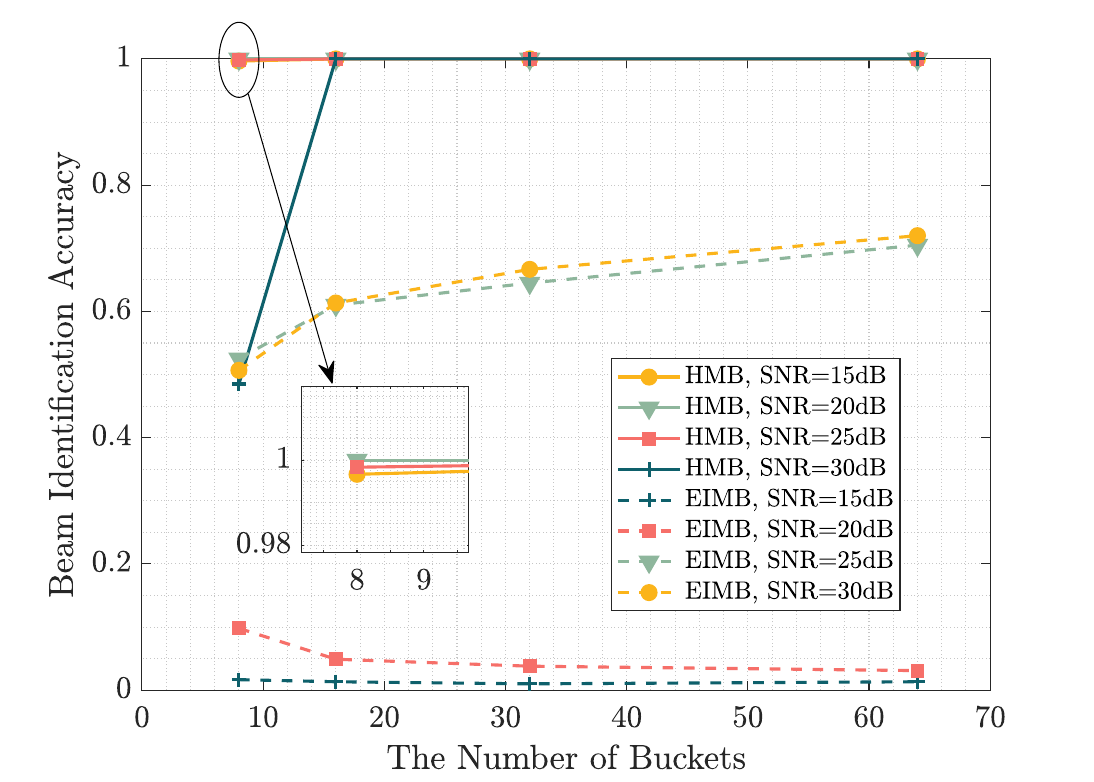}}  \vspace{-1mm}
			\caption{Success beam identification accuracy versus the number of buckets $B$ at different SNRs. }\vspace{-6mm}
			\label{accuracy-B-SNR} 
		\end{center}
\end{figure} 
\vspace{-9mm}
\section{Simulation Results}
\vspace{4mm}
We now evaluate the performance of our proposed beam training method with simulation results. The number of APs and the operating frequency are set to $K=5$ and $f_c=28\text{ GHz} $ respectively, and the signal wavelength is $\lambda_c=0.01\text{ m}$. The planar antenna arrays of APs contain $M=4$, $N=128$ antennas, and the spacing between the antennas is $d_x=d_z=\lambda_c/2 $. The reference SNR is $\gamma=\frac{P_0MN\rho_0}{r_0^2\sigma^2}$ with $\rho_0=-72$ dB, $P_0=15$ dBm, and $\sigma^2=-70$ dBm. The achievable rate, denoted in bits/second/Hz (bps/Hz), is given by
\begin{equation}
    R=\log_2(1+\gamma|\mathbf{f}_{BB}^T\mathbf{F}_{RF}^T\mathbf{g}_1|^2).
\end{equation}
\par
Fig.~\ref{fig:accuray} plots the effect of the SNR on the identification accuracy. We use the exhaustive, EIMB training method with the near-field codebook and exhaustive training with the DFT codebook ("Exhaustive-DFT") as the baseline. Firstly, it can be seen that as the SNR increases, the influence of noise diminishes, leading to improved identification accuracy for all beam training methods. Specifically, the accuracy achieved with the near-field codebook using the exhaustive beam training method converges to 1. However, the accuracy achieved with the far-field codebook is significantly lower, indicating the limitations of the DFT codebook in the near field.
\par
In addition, when utilizing near-field codebooks, it reveals that the HMB training method demonstrates competitive performance compared to the exhaustive training method. Specifically, when the number of multi-arm beams $B\geq 32$ and the SNR is not less than $5$ dB, the HMB training method can achieve at least $96.4$\% of the performance achieved by the exhaustive training method. Moreover, even at relatively low SNRs, the HMB training method exhibits considerably improved identification accuracy compared to the EIMB training method. 
\par
Fig.~\ref{accuracy-B-SNR} plots the effect of the number of buckets on the identification accuracy. It can be noticed that as the number of multi-arm beams $B$ decreases, the accuracy gradually decreases. The reason is that the number of sub-rays $R$ increases as $B$ decreases, making leakage interference among sub-beams more influential on identification results. It is also observed that a higher SNR is beneficial for beam training. Compared to the EIMB training method, the HMB training method shows an outstanding identification accuracy, which converges almost to $1$.
\begin{figure}[t]\vspace{-3mm}
	\begin{center}
			\centerline{\includegraphics[height= 0.382\textwidth]{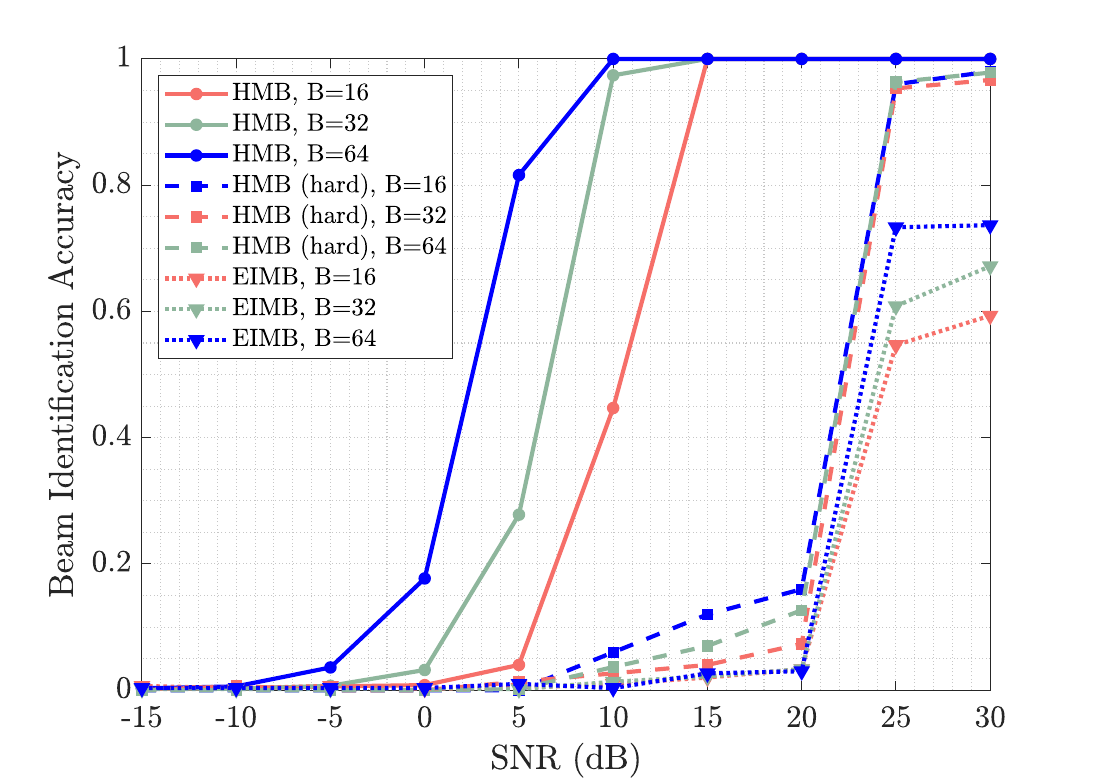}}  \vspace{-1mm}
			\caption{Success beam identification accuracy versus the SNR when considering soft and hard decisions.  }\vspace{-7mm}
			\label{fig:accuracy_softhard} 
		\end{center}
\end{figure} 
\par
Fig.~\ref{fig:accuracy_softhard} plots the beam identification accuracy versus the SNR when considering soft and hard decisions. Among the evaluated beam training methods, only the depicted "HMB" method utilizes the soft decision, while the "HMB (hard)" method employs the hard decision based on threshold comparison and the HMB codebook. It can be seen that our proposed HMB training method has the best performance, especially when the SNR is relatively small. Specifically, when the SNR is $10$ dB and the number of beams $B = 32$, the utilization of soft decision improves the accuracy by $96.9$\%. 
\par
This improvement is attributed to the fact that the use of multi-arm beams allocates the transmit power to each sub-beam, further reducing the power of the received signal, which is more likely to be drowned in the noise at low SNRs. Consequently, accurately and adaptively determining the threshold for the hard decision becomes challenging. The soft decision, on the other hand, compares relative values, eliminates the need for thresholding and is less affected by noise. In addition, when the SNR exceeds $20$ dB, the HMB codebook improves accuracy by $22$\% compared to the EIMB codebook. This is because the equal interval method has a fixed leakage interference, whereas the randomness of the hash adds a random perturbation to the leakage interference between sub-beams, attenuating the effect of this interference on the subsequent decision. 
\begin{figure}[t]
	\begin{center}
			\centerline{\includegraphics[height=   0.382\textwidth]{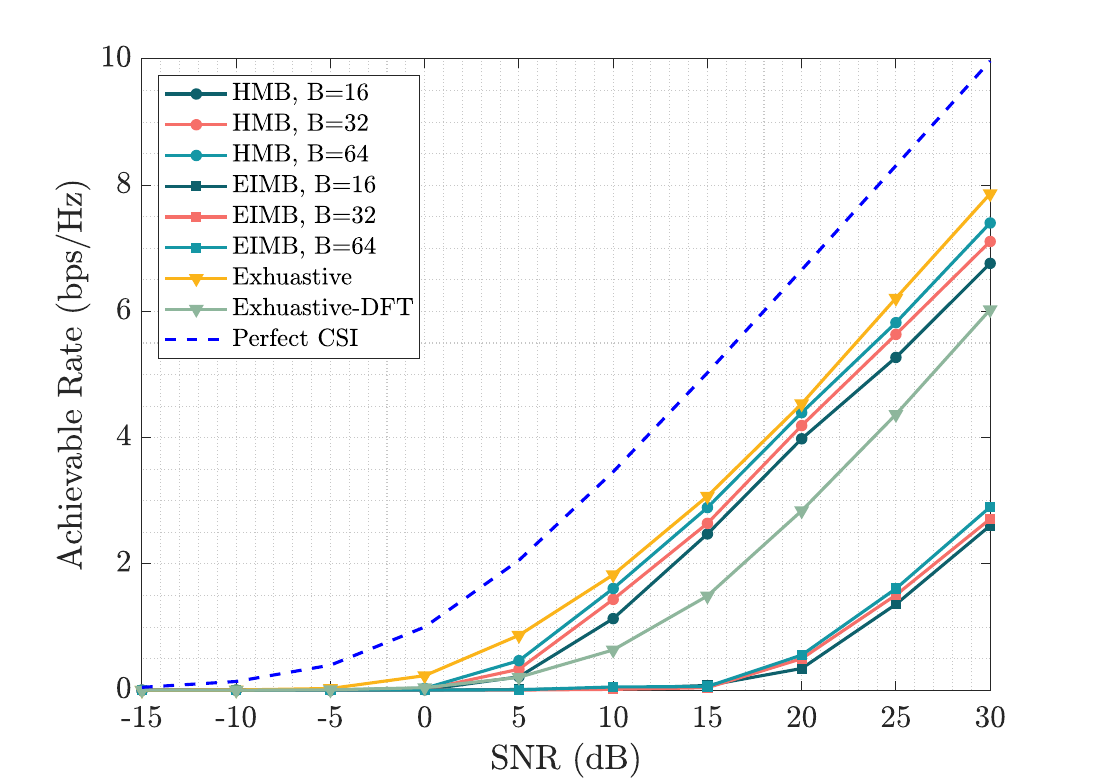}}  \vspace{-1mm}
			\caption{Achievable rate versus the SNR under different beam training schemes. }\vspace{-7mm}
			\label{fig:rate_SNR} 
		\end{center}
\end{figure} 
\begin{figure}[t]
	\begin{center}
			\centerline{\includegraphics[height=   0.382\textwidth]{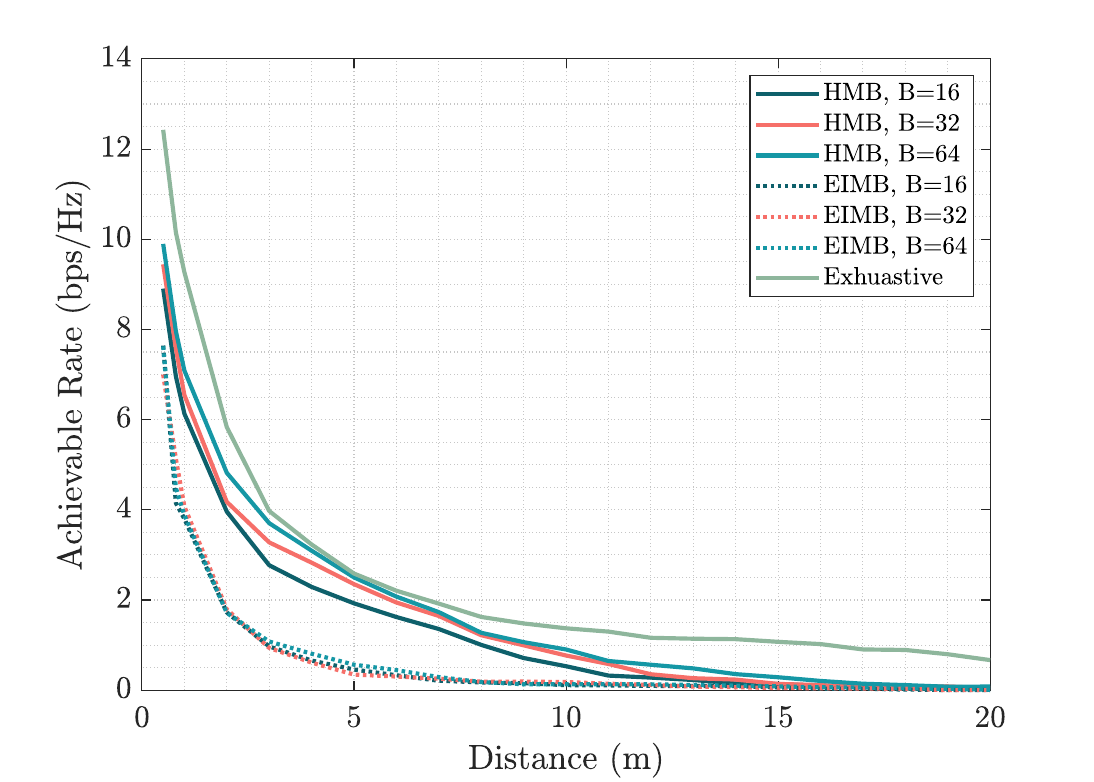}}  \vspace{-1mm}
			\caption{Achievable rate versus the distance between the AP and the user.  }\vspace{-7mm}
			\label{fig:rate_dist} 
		\end{center}
\end{figure} 

\begin{figure}[t]
	\begin{center}
	\centerline{\includegraphics[height=   0.382\textwidth]{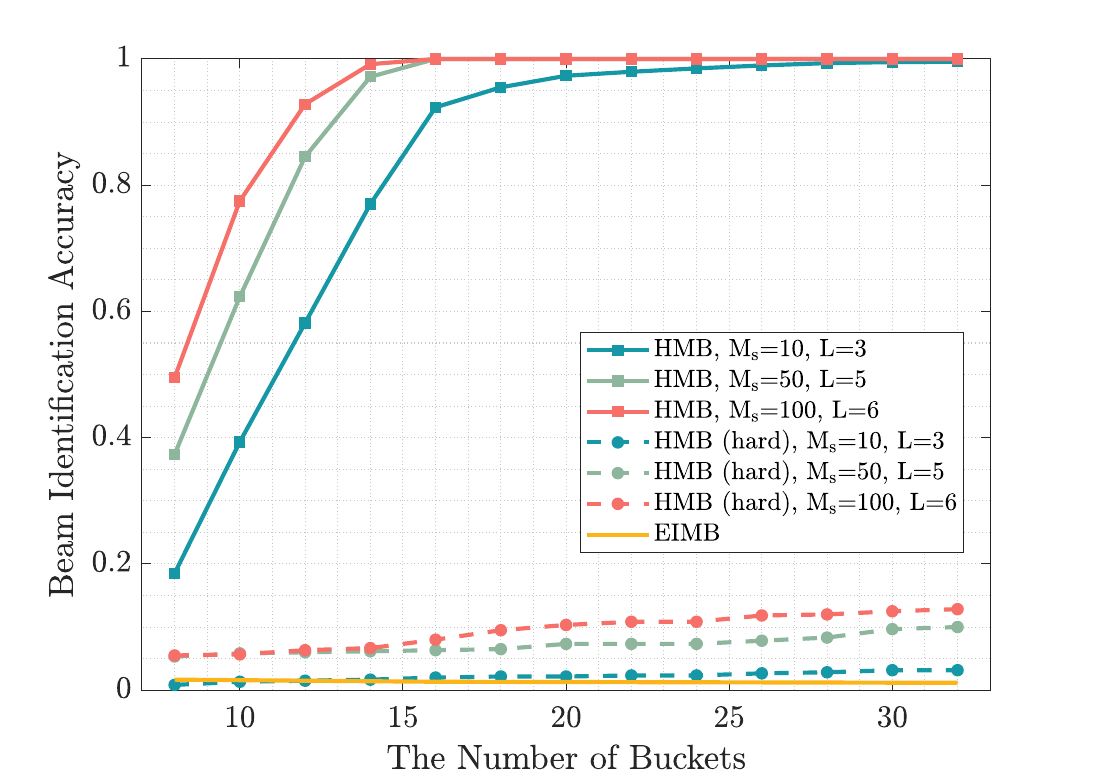}}  \vspace{-1mm}
			\caption{Success beam identification accuracy versus the number of buckets $B$ for different number of rounds.  }
			\label{accuracy-B-round} 
		\end{center}\vspace{-6.9mm}
\end{figure} 
\begin{figure}[t]
	\begin{center}
	\centerline{\includegraphics[height=   0.382\textwidth]{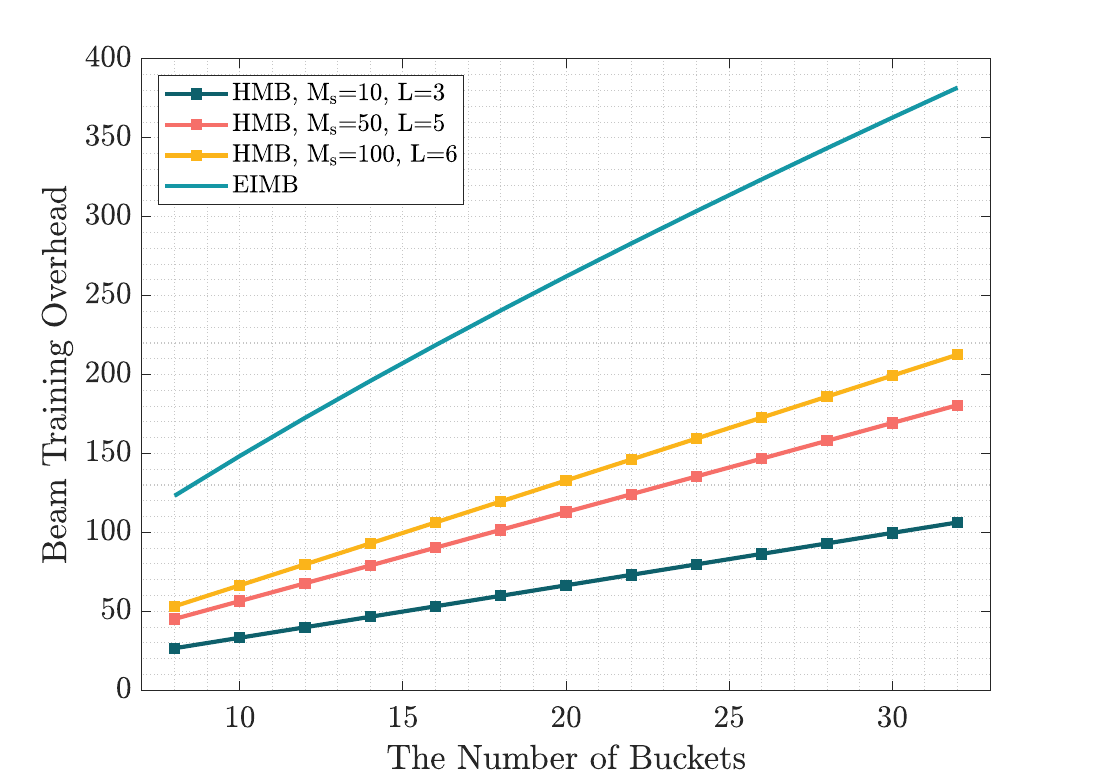}}  \vspace{-1mm}
			\caption{Training overhead versus the number of buckets $B$ for different number of rounds.}
			\label{overhead-B} 
		\end{center}\vspace{-6mm}
\end{figure} 
Fig.~\ref{fig:rate_SNR} plots the achievable rate of different beam training methods at different SNRs. It reveals that the achievable rate grows exponentially when the SNR is small, while the growth becomes almost linear as the SNR increases. Among the evaluated methods, the achievable rate of the exhaustive beam training method is closest to the performance of perfect CSI, and the HMB training method achieves $90$\% of the performance of the exhaustive beam training when $B = 64$. Similarly, it can be observed that both the exhaustive beam training with the DFT codebook and the EIMB training with the designed near-field codebook suffer severe performance losses.
\par
Fig.~\ref{fig:rate_dist} plots the effect of the AP-user distance on the achievable rate. It can be found that the achievable rate gradually decreases as the distance increases, and faster when the distance is smaller. In addition, the designed HMB training method significantly outperforms the EIMB training method and is infinitely close to the performance of the exhaustive training method.
\par
Fig.~\ref{accuracy-B-round} and Fig.~\ref{overhead-B} plot the effect of the number of hash rounds $L$ and buckets $B$ on training accuracy and overhead, respectively. There is a trade-off, a larger $L$ contributes to improved accuracy but introduces extra overhead. It can be found that when the number of buckets $B$ is smaller, it is more beneficial to increase the number of training rounds $L$. Specifically, when $B=14$, increasing $L$ from $3$ to $5$ improves the accuracy by 19.78\% while increasing training overhead by $28$ time slots, and when $B=32$ it improves accuracy by only 0.65\% while increasing complexity by $64$ time slots. Moreover, the accuracy achieved when $L=3$, with the corresponding $\frac{1}{M_s}=$10\%, is found to be nearly 90\% of the accuracy obtained when $L=6$ ($\frac{1}{M_s}=$1\%). This observation serves as a support for the derivation in section \ref{sec:complexity}.
\begin{figure}[t]
	\begin{center}
			\centerline{\includegraphics[height=   0.382\textwidth]{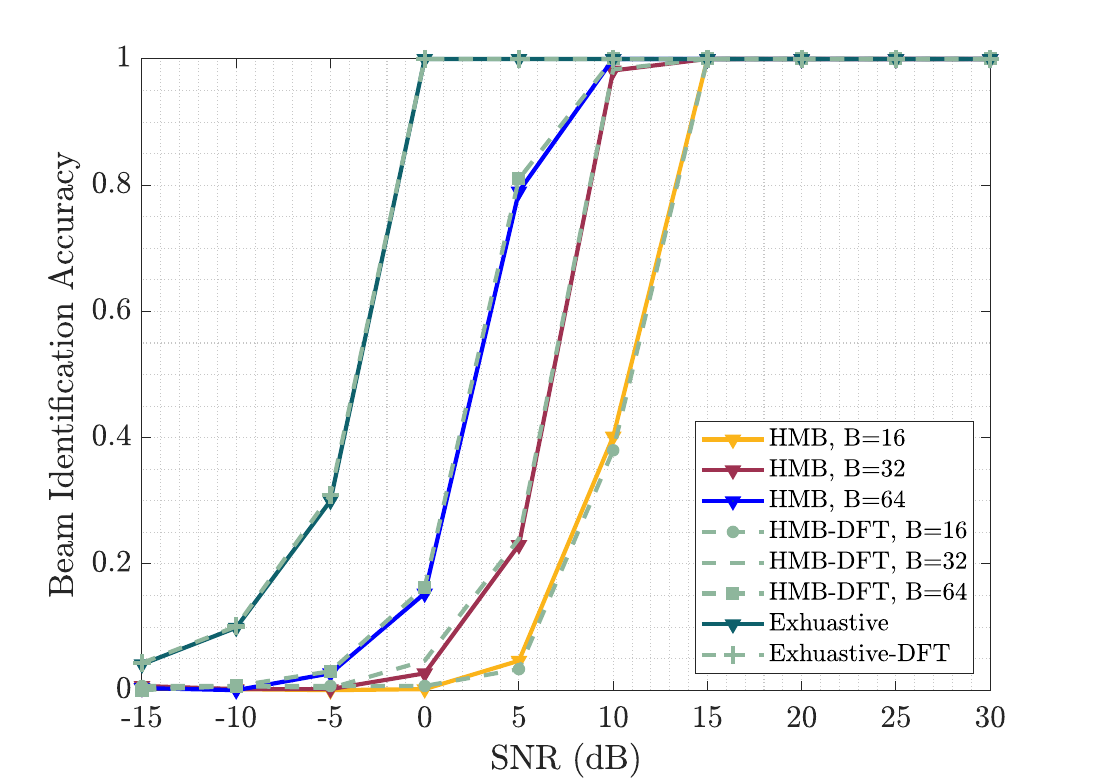}}  \vspace{-1mm}
			\caption{Success beam identification accuracy versus the SNR under far-field simulation condition.}
			\label{fig:accuracy_far} 
		\end{center}\vspace{-6.9mm}
\end{figure} 
\begin{figure}[t]
	\begin{center}
			\centerline{\includegraphics[height=   0.382\textwidth]{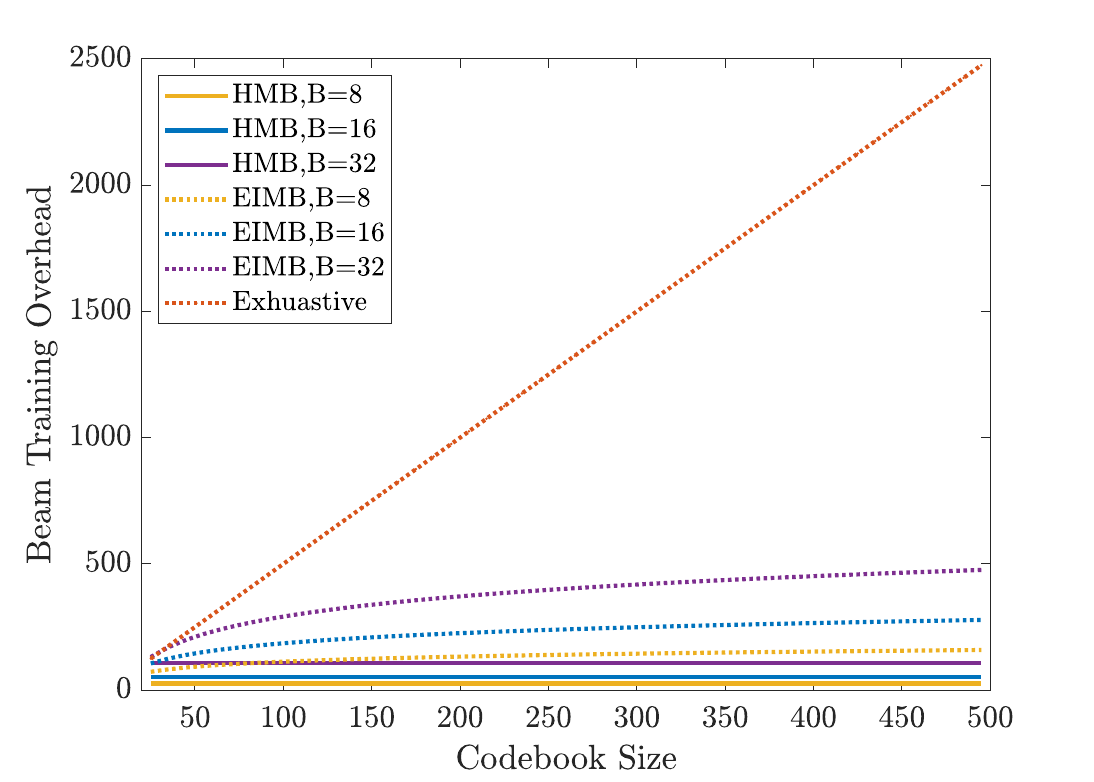}}  \vspace{-1mm}
			\caption{Training overhead versus codebook size under different beam training schemes.  }\vspace{-6mm}
			\label{fig:overhead} 
		\end{center}
\end{figure} 
\par
Fig.~\ref{fig:accuracy_far} plots the beam identification accuracy versus the SNR under far-field conditions. It can be seen that the codebook constructed using our proposed HMB training method achieves nearly identical accuracy compared to the DFT codebook, which verifies the applicability of the proposed HMB training method in the far-field region.
\par
\textcolor{black}{Fig.~\ref{fig:overhead} plots the effect of different codebook sizes on the beam training overhead. The exhaustive approach incurs a training time that is directly proportional to the codebook size and the number of APs $K$. The training overhead of the EIMB training method is also proportional to $K$. In contrast, our proposed HMB training method achieves a logarithmic level of training overhead. The training time for HMB is given by $Q=BL=O(B\mathrm{log}MN)$, which is independent of $K$ and significantly reduces the training overhead.}
\par
\vspace{-1.3mm}
\section{Conclusion}
\vspace{2mm}
\par
\textcolor{black}{In this paper, the HMB training method was proposed for IGAS networks. We take the near field as an instance and its applicability to the far field was verified. To begin with, the IGAS training single beams were constructed by exploiting the polar-domain sparsity. To further reduce the training overhead, independent hash functions were used to generate multi-arm beams. After all APs traversed the predefined beams simultaneously, the soft decision and voting mechanism were employed to improve the identification accuracy and obtain the best-aligned beam. Simulation results showed that our proposed beam training method maintained stable and satisfactory performance in terms of beam identification accuracy. It achieved an accuracy of 96.4\% compared to the exhaustive training method, while significantly reducing the training overhead to the logarithmic level.}
\vspace{0mm}
{\appendices

\section*{\textcolor{black}{APPENDIX A}} 
\section*{\textcolor{black}{Proof of Theorem \ref{theorem:k-wise}}}
\label{app:a}
\textcolor{black}{Define $a_0 + a_1x + a_2x^2 + ... + a_{k-1}x^{k-1}=r$, when we choose $k$ different values of $x\in GF(S)$, we obtain $k$ values of $r$, which can be written in the matrix form as}
\begin{equation}
    \begin{bmatrix}
        1 & x_1 &x_1^2&...&x_1^{k-1} \\
        1 & x_2 &x_2^2&...&x_2^{k-1} \\
        \vdots & \vdots & \vdots & \ddots   & \vdots  \\
        1 & x_k &x_k^2&...&x_k^{k-1}
    \end{bmatrix}
    \begin{bmatrix}
        a_0 \\
        a_1 \\
        a_2 \\
        \vdots\\
        a_{k-1}
    \end{bmatrix}
    =
    \begin{bmatrix}
        r_1 \\
        r_2 \\
        \vdots\\
        r_k
    \end{bmatrix}.
\end{equation}

The matrix containing $x$ is invertible because for $\forall i\neq j \in \{1,...,k\}$, we have $x_i\neq x_j$. Thus, after an inverse matrix operation over the finite field $GF(S)$, the coefficients of the hash function can be derived as
\begin{equation}
    \begin{bmatrix}
        a_0 \\
        a_1 \\
        a_2 \\
        \vdots\\
        a_{k-1}
    \end{bmatrix}
    =
    \begin{bmatrix}
        1 & x_1 &x_1^2&\cdots&x_1^{k-1} \\
        1 & x_2 &x_2^2&...&x_2^{k-1} \\
        \vdots & \vdots & \vdots & \ddots   & \vdots  \\
        1 & x_k &x_k^2&...&x_k^{k-1}
    \end{bmatrix}^{-1}
    \begin{bmatrix}
        r_1 \\
        r_2 \\
        \vdots\\
        r_k
    \end{bmatrix}.
\end{equation}
This implies that when fixing $x_1, x_2,...,x_k$, there is only a unique set of coefficients $a_0,a_1,...,a_{k-1}$ corresponding to $r_1,r_2,...,r_k$. The polynomial $a_0 + a_1x + a_2x^2 + ... + a_{k-1}x^{k-1}=r$ serves as a hash function, mapping the elements from the universe $\mathcal{U}$ to itself, and the hash values $r_1,r_2,...,r_k$ obtained from different sets of coefficients $a_0,a_1,...,a_{k-1}$ are independent of each other.

To form a family of hash functions that map $\mathcal{U}\to\mathcal{T}$, we discard $\lceil \log_2 S\rceil-\lceil \log_2B\rceil$ bits from the values $r_i$, $i=\{1,...,k\}$. It results in a collection of hash functions that effectively map elements from the universe $\mathcal{U}$ to the interpreted hash values within the designated range $\mathcal{T}=\{0, 1, ..., B-1\}$. That is,
\begin{align}\label{eq:pr1}
        &Pr(h(x_1)=\alpha_1\ \wedge\ h(x_2)=\alpha_2\ \wedge\ ...\ \wedge\ h(x_k)=\alpha_k)\nonumber\\
        &=Pr(r_1\ \rm{mod} \ B=\alpha_1\ \wedge\ r_2\ \rm{mod}\ B=\alpha_2\ \wedge\ ...\nonumber\\ 
        &\quad\wedge\ r_k\ \rm{mod}\ B=\alpha_k), \ \ \ \alpha_{i}\neq\alpha_{j}\in\mathcal{T}.
\end{align} 
\par
The randomly chosen coefficients $a_0, a_1, ..., a_{k-1}$ make the results of the polynomial follow the uniform distribution. Consequently, the outcomes obtained after applying the modulo operation exhibit the same probability distribution. That is, the probability that $r_i\ \rm{mod} \ B$ is equal to a given value $\alpha_i \in \mathcal{T}$ is at most $\frac{1}{B}$. Importantly, when $x_1\neq x_2\neq ...\neq x_k$, we obtain the desired property expressed by
\begin{equation}\label{eq:k-wise}
        Pr(h(x_1)=\alpha_1\wedge h(x_2)=\alpha_2\wedge...\wedge h(x_k)=\alpha_k)=\frac{1}{B^k},
    \end{equation}
which signifies the attainment of $k$-wise independence.

\section*{\textcolor{black}{APPENDIX B}} 
\section*{\textcolor{black}{Proof of Theorem \ref{theorem:complexity}}}
\begin{lemma}
(Hoeffding's Lemma) For random variables $X$, $P(X\in[a,b])=1$, $E[X]=0$, there are
\begin{equation}
	\begin{split}
		\forall s,\quad E[e^{sX}]\leq e^{(s^2 (b-a)^2/8)}.
	\end{split}
\end{equation}
\end{lemma}
\par
Firstly, for the $l\in\{1,...,L\}$-th round of hash, the power of the signal received by the $b$-th multi-arm beam is denoted as $P(l,b)$, where $q=(l-1)B+b$ represents the time slot number. Since the hash functions for the $L$ rounds of hash are randomly chosen from the hash function family, the values in $\mathcal{L}_b=\{P(1,b),...,P(l,b),...,P(L,b)\},\forall b$, are random variables. The number of elements in set $\mathcal{L}_b$ is denoted as $card(\mathcal{L}_b)=L_b$. Additionally, these variables are independent due to the $k$-wise independence property.  
\par
Furthermore, since the received signal power values are positive and finite, we have $P(l,b)\in[P_{\min},P_{\max}], l\in\{1,...,L\}$, where $P_{\min}$ and $P_{\max}$ represent the upper and lower bounds of the power, respectively. This leads to for $\forall s,t>0$,
\begin{align}\label{eq:Plb}
        &Pr(P(l,b)-E[P(l,b)]\geq t)=Pr(e^{s(P(l,b)-E[P(l,b)])} \geq e^{st})\nonumber\\
        &\overset{(b)}{\leq} e^{-st}E[e^{s(P(l,b)-E[P(l,b)])}]=e^{-st}\prod\limits_{l=1}^L E[e^{s\frac{P(l,b)-E[P(l,b)]}{L}}]\nonumber\\
        &\overset{(c)}{\leq} e^{-st}\prod\limits_{l=1}^L E[e^{s^2(\frac{P_{\max}-P_{\min}}{L})^2/8}]=e^{-st+\frac{s^2}{8}\sum\limits_{l=1}^{L}(\frac{P_{\max}-P_{\min}}{L})^2}\nonumber\\
        &=e^{-st+\frac{s^2(P_{\max}-P_{\min})^2}{8L}}.
\end{align}
The inequality $(b)$ is based on Markov's inequality because the random variable is non-negative and has a finite mean, which satisfies the preconditions of Markov's inequality, $(c)$ is based on Hoeffding's Lemma. Since the index $b$ can take the value from $1$ to $B$, and each multi-arm beam is equivalent due to the randomness of the hash function, we can use the same threshold $E[P(l,b)]\triangleq T_0 $. Then we have 
\begin{equation}
    \frac{P(l,b)-E[P(l,b)]}{L}\in[\frac{P_{\min}-T_0}{L},\frac{P_{\max}-T_0}{L}],
\end{equation}
\begin{equation}
    E[\frac{P(l,b)-E[P(l,b)]}{L}]=0.
\end{equation}
\par
To obtain the best probability upper bound, we minimize the right-hand side of the equation concerning $s$, define
\begin{equation}
    g(s)=-st+\frac{s^2 (P_{\max}-P_{\min} )^2}{8L}=-st+\frac{s^2(\Delta P)^2}{8L}.
\end{equation}
\par
This equation is a quadratic function that takes the minimum value when $s=\frac{4tL}{(\Delta P)^2}$, thus\vspace{-1mm}
\begin{equation}
    g(s)_{\min}=-\frac{2t^2L}{(\Delta P)^2},
\end{equation}
where $\Delta P\triangleq  P_{\max}-P_{\min}$.
\par\vspace{2mm}
Consequently, we have\vspace{-1mm}
\begin{equation}
	\begin{split}\label{eq:eq1}
		\forall t>0,\forall b,\quad Pr(P(l,b)-E[P(l,b)]\geq t) \leq e^{-\frac{2t^2L}{(\Delta P)^2}}.
	\end{split}
\end{equation}
\par\vspace{1mm}
Because $P(l,b)-E[P(l,b)]=-(E[P(l,b)]-P(l,b))$, in the same way we can obtain
\vspace{-2mm}
\begin{equation}
	\begin{split}\label{eq:eq2}
		\forall t>0,\forall b,\quad Pr(E[P(l,b)]-P(l,b)\geq t) \leq e^{-\frac{2t^2L}{(\Delta P)^2}}.
	\end{split}
\end{equation}
\par\vspace{1mm}
Divide the random variable $P(1,b),...,P(l,b),...,P(L,b)$ into those that contain user signals and those that do not (i.e., only the noise), denote as $\mathcal{L}_b^{s}=\{P(l,b)^{(s)}\}$ and $\mathcal{L}_b^{ns}\{P(l,b)^{(ns)}\}$, respectively. We have\vspace{1mm} 
\begin{equation}
    \mathcal{L}_b^{s}\in[P_{\min}^{s},P_{\max}^{s}],\quad E[\mathcal{L}_b^{s}]\triangleq T_b^{s},
\end{equation}
\begin{equation}
    \mathcal{L}_b^{ns}\in[P_{\min}^{ns},P_{\max}^{ns}],\quad E[\mathcal{L}_b^{ns}]\triangleq T_b^{ns},\vspace{1mm}
\end{equation}
where $P_{\min}^{s},P_{\max}^{s}$ are the lower and upper bounds of the receive power of the useful signal, $P_{\min}^{ns},P_{\max}^{ns}$ are the lower and upper bounds of the received power of the noise. \vspace{1mm}
\begin{equation}
\begin{split}
    card(\mathcal{L}_b^{s})&=L_b^{s},\quad card(\mathcal{L}_b^{ns})=L_b^{ns},\\
    &L_b^{s}+L_b^{ns}=L_b.
\end{split}
\end{equation}
\par
Since 
\begin{equation}
    T_b^{s} L_b^{s}+T_b^{ns} L_b^{ns}=T_0L_b,
\end{equation}
we have $T_b^{ns}<T_0<T_b^{s}$. 
\par
According to (\ref{eq:eq1}), if we take $t=T_0-T_b^{ns}>0$ and 
\begin{equation}
    L^{ns}=\frac{\mathrm{ln}M_s (P_{\max}^{ns}-P_{\min}^{ns})^2}{2(T_0-T_b^{ns})^2},
\end{equation}
we have \vspace{-2mm}
\begin{equation}
	\begin{split}
		Pr(P(l,b)^{(ns)}\geq T_0 )\leq e^{-\frac{2(T_0-T_b^{ns})^2 L^{ns}}{(P_{\max}^{ns}-P_{\min}^{ns})^2 }}=\frac{1}{M_s}.
	\end{split}
\end{equation}
\par
Similarly, according to (\ref{eq:eq2}), if we take $t=T_b^{s}-T_0>0$ and
\begin{equation}
    L^{s}=\frac{\mathrm{ln}M_s (P_{\max}^{s}-P_{\min}^{s})^2}{2(T_0-T_b^{s})^2},
\end{equation}
we have \vspace{-2mm}
\begin{equation}
	\begin{split}
		Pr(P(l,b)^{(s)}\leq T_0 )\leq e^{-\frac{2(T_0-T_b^{s})^2 L^{s}}{(P_{\max}^{s}-P_{\min}^{s})^2 }}=\frac{1}{M_s},
	\end{split}
\end{equation}
\begin{equation}\label{eq:L}
    L^{ns}=O(\mathrm{log}M_s),\quad L^{s}=O(\mathrm{log}M_s).
\end{equation}
\par
Eventually, the probability of identification error is reduced to $\frac{1}{M_s}$ when (\ref{eq:L}) holds, we obtain
\begin{equation}
    L=L^s+L^{ns}=O(\mathrm{log}M_s).
\end{equation}
}
\bibliographystyle{IEEEbib}
\bibliography{myrefs}
\end{document}